\documentclass[]{aastex631}
\usepackage{amsmath}
\usepackage{xcolor}
\usepackage{CJK}
\usepackage{multirow}
\usepackage{tabularx}

\usepackage{dblfloatfix}

\usepackage{chngcntr}

\submitjournal{ApJL}
\received{Sept 10, 2024}
\revised{Dec 3, 2024}
\accepted{Dec 8, 2024}

\begin{document}
\title{Volatile-rich Sub-Neptunes as Hydrothermal Worlds: The Case of K2-18 b}

\correspondingauthor{Xinting Yu}
\email{xinting.yu@utsa.edu}

\author[0000-0001-5486-1330]{Cindy N. Luu}
\affiliation{Department of Physics and Astronomy, University of Texas at San Antonio\\
1 UTSA Circle, San Antonio, TX 78249, USA}

\author[0000-0002-7479-1437]{Xinting Yu\begin{CJK*}{UTF8}{gbsn}
(余馨婷)\end{CJK*}}
\affiliation{Department of Physics and Astronomy, University of Texas at San Antonio\\
1 UTSA Circle, San Antonio, TX 78249, USA}

\author[0000-0002-2161-4672]{Christopher R. Glein}
\affiliation{Space Science Division, Southwest Research Institute\\
6220 Culebra Rd, San Antonio, TX 78238, USA}

\author[0000-0001-5271-0635]{Hamish Innes}
\affiliation{Department of Earth Sciences, Freie Universit{\"a}t Berlin\\Malteserstr. 74-100, 12249 Berlin, Germany}
\affiliation{Institute of Planetary Research, German Aerospace Center (DLR)\\ Rutherfordstra{\ss}e 2, 12489 Berlin, Germany}

\author[0000-0002-8949-5956]{Artyom Aguichine}
\affiliation{Department of Astronomy and Astrophysics, University of California Santa Cruz\\
1156 High Street, Santa Cruz, California 95064, USA}

\author[0000-0001-6878-4866]{Joshua Krissansen-Totton}
\affiliation{Department of Earth and Space Sciences, University of Washington, Seattle\\
1410 NE Campus Pkwy, Seattle, WA 98195, USA}

\author[0000-0002-8837-0035]{Julianne I. Moses}
\affiliation{Space Science Institute, \\
Boulder, Colorado 80301, USA}

\author[0000-0002-8163-4608]{Shang-Min Tsai}
\affiliation{Department of Earth and Planetary Sciences, University of California Riverside\\
900 University Ave, Riverside, CA 92521, USA}

\author[0000-0002-8706-6963]{Xi Zhang}
\affiliation{Department of Earth and Planetary Sciences, University of California Santa Cruz\\
1156 High St, Santa Cruz, CA 95064, USA}

\author[0000-0003-2689-3102]{Ngoc Truong}
\affiliation{Space Science Division, Southwest Research Institute\\
6220 Culebra Rd, San Antonio, TX 78238, USA}

\author[0000-0002-9843-4354]{Jonathan J. Fortney}
\affiliation{Department of Astronomy and Astrophysics, University of California Santa Cruz\\
1156 High Street, Santa Cruz, California 95064, USA}

\begin{abstract}

Temperate exoplanets between the sizes of Earth and Neptune, known as ``sub-Neptunes'', have emerged as intriguing targets for astrobiology. It is unknown whether these planets resemble Earth-like terrestrial worlds with a habitable surface, Neptune-like giant planets with deep atmospheres and no habitable surface, or something exotic in between. Recent JWST transmission spectroscopy observations of the canonical sub-Neptune, K2-18 b, revealed $\sim$ 1\% CH$_4$, $\sim$ 1\% CO$_2$, and a non-detection of CO in the atmosphere. While previous studies proposed that the observed atmospheric composition could help constrain the lower atmosphere's conditions and determine the interior structure of sub-Neptunes like K2-18 b, the possible interactions between the atmosphere and a hot, supercritical water ocean at its base remain unexplored. In this work, we investigate whether a global supercritical water ocean, resembling a planetary-scale hydrothermal system, can explain these observations on K2-18 b-like sub-Neptunes through equilibrium aqueous geochemical calculations. We find that the observed atmospheric CH$_4$/CO$_2$ ratio implies a minimum ocean temperature of $\sim710~\rm K$, whereas the corresponding CO/CO$_2$ ratio allows ocean temperatures up to $\sim 1070~\rm K$. These results indicate that a global supercritical water ocean on K2-18 b is plausible. While life cannot survive in such an ocean, this work represents the first step towards understanding how a global supercritical water ocean may influence observable atmospheric characteristics on volatile-rich sub-Neptunes. Future observations with better constrained CO and NH$_3$ mixing ratios could further help distinguish between possible interior compositions of K2-18 b.
\end{abstract}

\keywords{\href{http://astrothesaurus.org/uat/498}{Exoplanets (498)}, \href{http://astrothesaurus.org/uat/487}{Exoplanet atmospheres (487)}, \href{http://astrothesaurus.org/uat/2310}{Exoplanet atmospheric structure (2310)}, \href{http://astrothesaurus.org/uat/1244}{Planetary atmospheres (1244)}, \href{http://astrothesaurus.org/uat/74}{Astrobiology (74)},
\href{http://astrothesaurus.org/uat/1151}{Ocean planets (1151)},
\href{http://astrothesaurus.org/uat/2291}{James Webb Space Telescope (2291)}
\href{http://astrothesaurus.org/uat/2172}{Extrasolar gaseous planets (2172)}}

\section{Introduction}
Understanding the compositions of exoplanets is crucial for identifying potentially habitable worlds. The majority of confirmed exoplanets are classified as sub-Neptunes -- exoplanets between the sizes of Earth and Neptune -- that are unlike any planetary body in our solar system \citep{Fressin_2013, Batalha_2014, Winn_2015, Fulton_2017}. Because a range of interior compositions can explain the observed masses and radii of sub-Neptunes, it is unknown whether these planets are closer to (1) terrestrial planets with thin atmospheres and potentially habitable surfaces, (2)``mini-Neptunes'' which are less massive versions of our solar system's ice giants, or (3) ``in-between state'' planets, such as ``water worlds" with volatile-rich interiors. Among the sub-Neptune population, a canonical habitable-zone sub-Neptune K2-18 b has captured the attention of the astrobiology community \citep{Glein_2024}.  

\begin{figure}[ht]
\centering
\includegraphics[width=\linewidth]{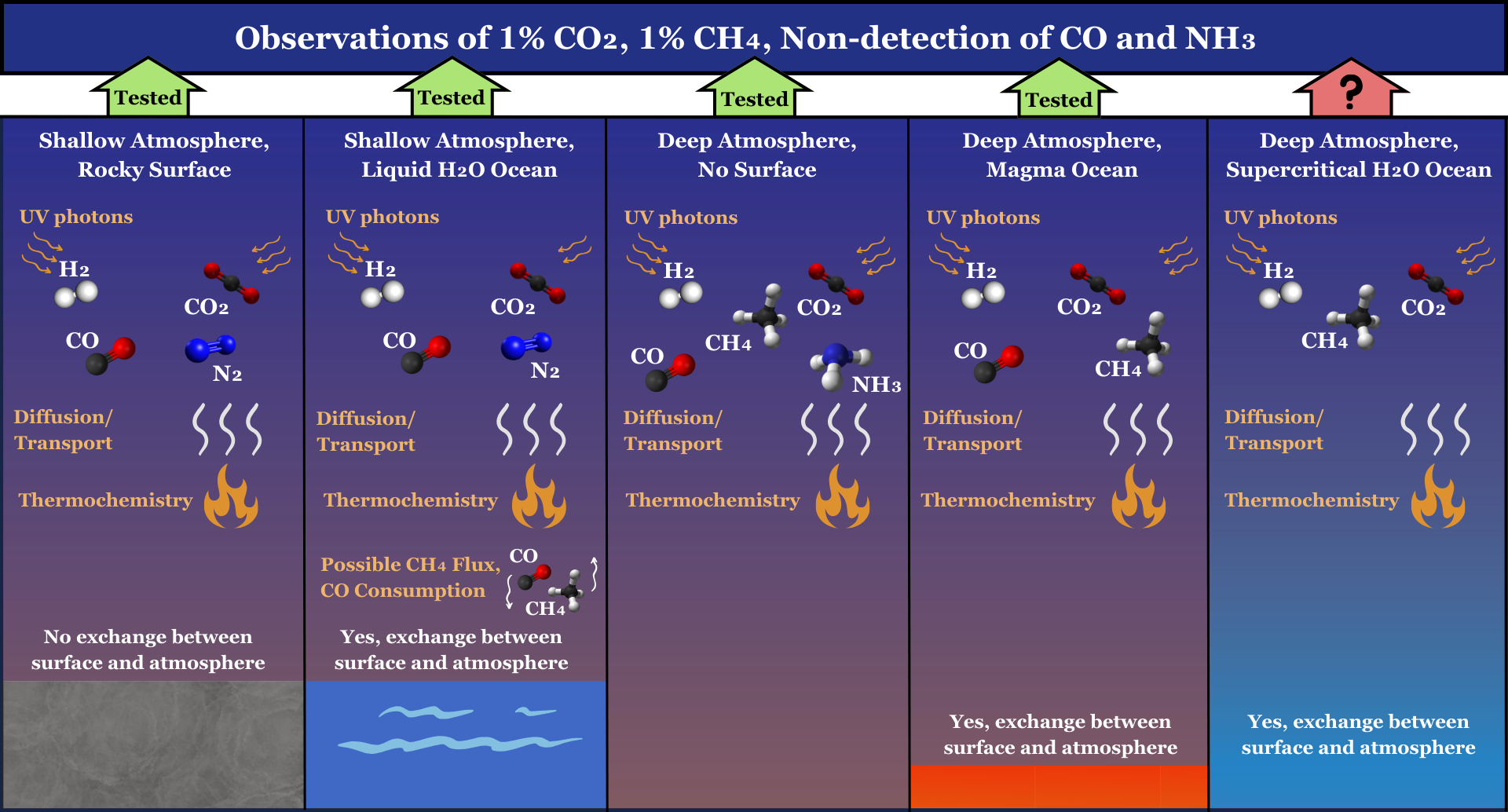}
\caption{Possible conceptual structures for K2-18 b that are explored previously \citep{Hu_2021_Photochemistry, Hu_2021_Thermochemistry, Tsai_2021, Yu_2021, Madhu_Hycean_2023,  Shorttle_2024, wogan2024, Rigby_2024, Cooke_2024} and in this work. 
\label{fig:visual}}
\end{figure}

K2-18 b has a radius of 2.610 $\pm$ 0.087 $R_{\oplus}$ \citep{Benneke_2019}, a mass of 8.63 $\pm$ 1.35 $M_{\oplus}$ \citep{Cloutier_2019}, and a bulk density of $2.67^{+0.52}_{-0.47}$ g/cm$^3$ \citep{Benneke_2019} that can be explained by a diverse range of interior structures with varying amounts of H$_2$O and H$_2$/He in a volatile-rich envelope \citep{Madhusudhan2020, Nixon_2021, Rigby_2024}. To discern the most plausible structure, previous studies proposed that the observed atmospheric composition could be key to understanding the conditions at the base of the atmosphere in temperate sub-Neptunes \citep{Hu_2021_Thermochemistry, Tsai_2021, Yu_2021, Madhu_Faraday_2023}. Potential atmospheric compositions have been explored for various possible atmosphere-interior structure scenarios (as shown in Figure~\ref{fig:visual}), including a thin-atmosphere rocky world \citep{Tsai_2021, Yu_2021}, a thin-atmosphere liquid water world with potential signs of life (known as a ``Hycean" world, \citealt{Hu_2021_Thermochemistry}; \citealt{Madhu_Hycean_2023}; \citealt{wogan2024}; \citealt{Tsai_2024}; \citealt{Cooke_2024}), a deep-atmosphere mini-Neptune with no surface \citep{Hu_2021_Photochemistry, Tsai_2021, Yu_2021, wogan2024}, and a deep-atmosphere mini-Neptune with a silicate magma ocean \citep{Shorttle_2024}. While the rocky world scenario was found to be inconsistent with observations \citep{Madhusudhan2020}, the water world and mini-Neptune scenarios remain possible.

Recent JWST transmission spectroscopy of K2-18 b led to the first detection of CH$_4$ ($\sim$ 1\%) and CO$_2$ ($\sim$ 1\%) in the atmosphere of a sub-Neptune, along with nondetections of H$_2$O, CO, and NH$_3$ \citep{Madhu_Hycean_2023}.  Despite all of the excellent work that has been done, existing models have struggled to fully explain this observed atmospheric composition. Recent predictions suggest K2-18 b's interior may be water-rich \citep{Yang_2024}, and sub-Neptunes with close orbits could host supercritical water oceans \citep{Mousis_2020}. Climate models also predict that ``Hycean worlds" (T$_\text{eq}$ $<$ 500 K, \cite{Madhu_2021}) receiving solar-like flux with pure H$_2$-He atmospheres will experience runaway greenhouse effects, driving surface water to a supercritical state \citep{Innes_2023, Pierrehumbert_2023, Leconte_2024}. However, the atmospheric interactions with a global supercritical water ocean remain largely unexplored. Pure water reaches its critical point at 220.64 bar and 647.096 K \citep{Wagner_2002}, beyond which there is no distinction between liquid and gas. On K2-18 b, an ocean of supercritical water would manifest as a layer of hot aqueous fluids merging seamlessly with the H$_2$-dominated atmosphere above, potentially creating a global hydrothermal environment. In this regime, we consider two distinct structures containing supercritical water to be likely. At lower pressures, H$_2$ is miscible with water \citep{Seward_1981, Innes_2023, Pierrehumbert_2023, Gupta_2024} and creates a homogeneous supercritical H$_2$-H$_2$O ocean. At higher pressures, H$_2$ and H$_2$O may be immiscible and could separate into a distinct supercritical H$_2$O-rich layer beneath a supercritical H$_2$-rich layer \citep{Gupta_2024}. Current photochemical models do not account for interactions between an H$_2$ atmosphere and a supercritical water or hydrogen layer \citep{Hu_2021_Photochemistry, Tsai_2021, Yu_2021, Madhu_Faraday_2023, Madhu_Hycean_2023}, highlighting the need for new approaches. Our study addresses this gap by exploring the geochemistry in a supercritical water ocean on K2-18 b in an attempt to explain the observed abundances of key species in K2-18 b's atmosphere. We focus on major carbon species as their abundances are well constrained by existing observations \citep{Madhu_Hycean_2023}. 


\section{Methodology}
We aim to derive constraints on the boundary conditions at the transition region between the atmosphere and a hypothetical supercritical water ocean. To explore how ocean conditions could drive volatile chemistry, we first calculate the aqueous equilibrium ratios of CH$_4$/CO$_2$ and CO/CO$_2$ in the supercritical water ocean, as detailed in Section~\ref{sub:aqueous}. Given the likely absence of a distinct boundary between the supercritical water ocean and the gaseous atmosphere, we conduct the aqueous calculations at temperatures and pressures where chemical equilibrium could be achieved. Consequently, the term ``ocean temperature" in this study refers to the condition for reaching an apparent chemical equilibrium rather than a discrete physical boundary between the ocean and the atmosphere.

Building on the aqueous results, we then explore how the aqueous ratios translate to the observable atmosphere. To achieve this, we performed exsolution calculations to predict the gaseous ratios of carbon-bearing species in the atmosphere, as detailed in Section \ref{sub:gaseous} (see also Figure \ref{fig:figure2}b). By comparing the predicted gaseous ratios to the observed species abundances in the atmosphere, we can determine whether a set of conditions in the supercritical water ocean would satisfy observational constraints.

\subsection{Aqueous Chemistry of Carbon}
\label{sub:aqueous}
At chemical equilibrium, the molar ratio of CH$_4$/CO$_2$ can be computed as a function of temperature, total pressure, and fluid oxidation state. In an H$_2$-rich environment such as K2-18 b, the oxidation state can be effectively represented by the molal concentration of hydrogen gas dissolved in the ocean. The reaction that describes the chemical equilibrium between CO$_2$ and CH$_4$ in an H$_2$-bearing supercritical fluid can be written as,

\begin{gather}
    \text{CO}_2\text{(aq)} + 4\text{H}_2\text{(aq)} \leftrightarrow \text{CH}_4\text{(aq)} + 2\text{H}_2\text{O(sc)}\label{eq:reaction1} \\
    {K}_1 = \left(\frac{{a}_{\text{CH}_4}}{{a}_{\text{CO}_2}}\right) 
    \frac{({a}_{\text{H}_2\text{O}})^2}{(a_{\text{H}_2})^4}
    \approx
    \left(\frac{{m}_{\text{CH}_4}}{{m}_{\text{CO}_2}}\right)\frac{({X}_{\text{H}_2\text{O}})^2}{(m_{\text{H}_2})^4}
    \label{eq:equilibrium1}
\end{gather}

where ``sc" refers to the supercritical fluid phase, $K$ stands for the equilibrium constant, $a$ denotes the activity in an aqueous solution, $X$ represents the mole fraction and $m$ designates the molality, which describes the number of moles of solute per kilogram of H$_2$O. In Equation \ref{eq:equilibrium1}, following \citet{Glein_2008}, we make the following approximations: 1) the activity of water is considered equivalent to the mole fraction of water in the supercritical phase; 2) the activity of H$_2$ is approximated by the molality of H$_2$, and 3) the activity ratio of CO$_2$ and CH$_4$ is equivalent to their molality ratio, and thus, their mole fraction ratio (${X}_{\text{CH}_4}/{X}_{\text{CO}_2}$). Other approximations for water and the aqueous species that reflect thermodynamic standard state conventions can be found in Appendix~\ref{sec:assumptions}. 

To evaluate for internal consistency, the ratio of CO/CO$_2$ can be estimated using the observed CH$_4$/CO$_2$ ratio constraint from \citet{Madhu_Hycean_2023}, based on the following reaction:

\begin{gather}
    \text{CH}_4\text{(aq)} + 4\text{CO}_2\text{(aq)} \leftrightarrow \text{CO}_2\text{(aq)} + 4\text{CO}\text{(aq)} +2\text{H}_2\text{O(sc)} 
    \label{eq:reaction2} \\
    {K}_2 = \left(\frac{a_{\text{CO}_2}}{a_{\text{CH}_4}}\right)
    \left(\frac{a_{\text{CO}}}{a_{\text{CO}_2}}\right)^4({a}_{\text{H}_2\text{O}})^2
    \approx 
    \left(\frac{m_{\text{CO}_2}}{m_{\text{CH}_4}}\right)
    \left(\frac{m_{\text{CO}}}{m_{\text{CO}_2}}\right)^4({X}_{\text{H}_2\text{O}})^2
    \label{eq:equilibrium2}
\end{gather} 

We chose this reaction because the observed mixing ratios of CH$_4$ and CO$_2$ make them the most abundant oxygen and carbon-bearing molecules and they are also the most well-constrained, while the uncertainty on CO abundance is high. By including CO$_2$ as both a product and reactant, we can express Equation~\ref{eq:equilibrium2} in terms of the carbon species ratios. This choice ensures that our CO/CO$_2$ modeling always aligns with the observed CH$_4$/CO$_2$ ratio. 

Here we implement the Deep Earth Water (DEW) model, originally developed to study the carbon cycle in Earth's mantle fluids \citep{Shock_1988, SHOCK_HELGESON_1990, SVERJENSKY2014125}, to calculate equilibrium constants under conditions potentially relevant to K2-18 b and determine the molal ratios of CH$_4$/CO$_2$ and CO/CO$_2$ across wide ranges of temperatures, pressures, and fluid oxidation states. Below, we consider ``ocean temperatures" ranging from 374$^{\circ}$C up to 1000$^{\circ}$C (647.15 K - 1273.15 K), where we reference 647.15 K as the critical temperature of pure water, and three pressure conditions (1 kbar, 5 kbar, and 10 kbar) to understand how key geochemical variables affect aqueous ratios involving CO$_2$, CH$_4$, and CO at chemical equilibrium. 

\subsection{Constraining the Range of Hydrogen Molality}\label{subsec:molality}

To model the CH$_4$/CO$_2$ ratios using Equation \ref{eq:equilibrium1}, we require an estimate of the hydrogen molality ($m_{\text{H}_2}$, expressed in molal units, mol$\cdot$(kg H$_2$O)$^{-1}$), based on the aqueous standard state on K2-18 b. Previous interior structure modeling by \citet{Madhusudhan2020} suggests that the overall $m_{\text{H}_2}$ of K2-18 b varies widely depending on the planet's bulk composition, from near-zero for water-dominated interiors to higher values for H-He dominated cases. Based on these models, combined with planet formation constraints, we estimate $m_{\text{H}_2}$ to plausibly range from 0.14 to 56.55 mol$\cdot$(kg H$_2$O)$^{-1}$, although higher values cannot be excluded. For further details on how these limits were derived, refer to Appendix~\ref{sec:H2_molality}.

However, the modeling results using $m_{\text{H}_2}$ equal and above the upper limit should only be interpreted qualitatively in terms of trends, as higher hydrogen molalities describe increasingly water-poor solutions. These conditions become less applicable to a water-rich supercritical ocean, which is the scenario that can be evaluated based on the assumptions inherent to our aqueous chemistry framework (See Appendix~\ref{sec:assumptions}). As such, an intermediate hydrogen molality of $m_{\text{H}_2} = 6.16$ mol$\cdot$(kg H$_2$O)$^{-1}$ was also selected to represent a mole fraction of 90.0\% H$_2$O and 10\% H$_2$, excluding He in this case. This value is roughly the upper limit that the DEW model can accommodate while preserving the dilute solution assumptions made in the model.

These hydrogen molalities, $m_{\text{H}_2}$ = 0.14, 6.16, and 56.55 mol$\cdot$(kg H$_2$O)$^{-1}$, correspond to hydrogen mole fractions of $X_{\text{H}_2}$ = 0.002, 0.100, and 0.433 and water mole fractions of $X_{\text{H}_2\text{O}}$ = 0.996, 0.900, and 0.424, respectively. For the remainder of this paper, we refer to these compositions as $X_{\text{H}_2\text{O}}$ = 99.6\%, 90.0\%, and 42.4\%.

\subsection{Gaseous Carbon Species from Ocean to Atmosphere}\label{sub:gaseous}
While understanding the carbon ratios in the ocean is an important first step, it is also important to understand how those ratios would be translated to the upper atmosphere where atmospheric observations are made. K2-18 b presents a special case due to water condensation, which depletes water vapor in the upper atmosphere and creates a distinct cloud layer of liquid water beneath the observable photosphere \citep{Benneke_2019, Madhu_Faraday_2023}. Hence, we must apply Henry's Law to predict the ratios of atmospheric gases involving CH$_4$, CO$_2$, and CO.

Henry's Law states that the activity of a solute species in a solvent is proportional to the fugacity of the gas in equilibrium with the solution at constant temperature and pressure. In ideal fluids, the fugacity is equivalent to the partial pressure. In this context, water serves as the solvent and gaseous species (CH$_4$, CO$_2$, and CO) dissolve in it. Here, we apply Henry's Law at the saturation boundary between the liquid and gaseous phases to estimate gas-phase ratios at the base of the cloud layer, where phase partitioning occurs.

The relevant reactions and equations corresponding to Henry's Law are shown below.

\begin{gather}
    \text{CH}_4\text{(g)}  \leftrightarrow \text{CH}_4\text{(aq)} \label{eq:reaction3} \\
    {K}_{\text{H,CH}_4} \approx \frac{(m_{\text{CH}_4})}{(p_{\text{CH}_4})} 
    \label{eq:henrys1}
\end{gather}

\begin{figure}[ht]
\centering
\includegraphics[width=1\linewidth]{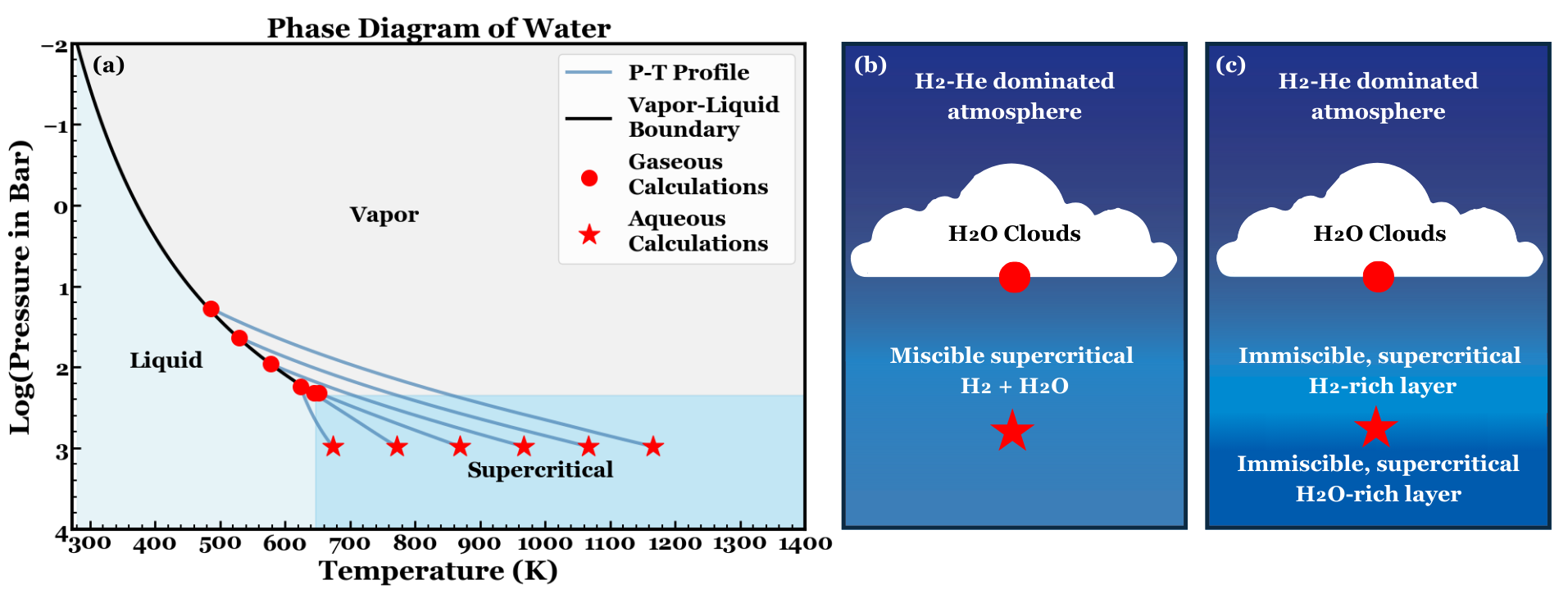}
\caption{(a) Phase diagram of pure water, showing adiabats for six selected ocean temperatures from 647.15 K to 1273.15 K at a pressure of redox equilibrium of 1 kbar as an example. This plot is most relevant to cases with lower H$_2$ abundances in the fluid envelope. Liquid water will be saturated at lower temperatures for the highest H$_2$ abundance considered in this work (see Appendix). (b) A schematic diagram showing the upper interior structure of K2-18 b with a miscible supercritical H$_2$+H$_2$O layer \citep{benneke2024jwst}, or (c) with phase-separated H$_2$-rich and H$_2$O-rich layers and a cloud layer in an H$_2$-He dominated atmosphere. Star and circle symbols represent where aqueous and gaseous calculations are performed, respectively.
\label{fig:figure2}}
\end{figure}

\begin{gather}    
    \text{CO}_2\text{(g)}  \leftrightarrow \text{CO}_2\text{(aq)} \label{eq:reaction4} \\
    {K}_{\text{H,CO}_2} \approx \frac{(m_{\text{CO}_2})}{(p_{\text{CO}_2})} \label{eq:henrys2}\\
    \text{CO}\text{(g)}  \leftrightarrow \text{CO}\text{(aq)} \label{eq:reaction5} \\
    {K}_{\text{H,CO}} \approx \frac{(m_{\text{CO}})}{(p_{\text{CO}})} 
    \label{eq:henrys3} 
\end{gather}

where ${K}_\text{H}$ represents the Henry's Law constant and $p$ corresponds to the partial pressure of each species in bars. Using these constants and the aqueous ratios, the gaseous ratio between CH$_4$/CO$_2$ and CO/CO$_2$ can be calculated as:

\begin{gather}
    \frac{p_{\text{CH}_4}}{p_{\text{CO}_2}} = \left(\frac{{K}_{\text{H,CO}_2}}{{K}_{\text{H,CH}_4}}\right)\left(\frac{m_{\text{CH}_4}}{m_{\text{CO}_2}}\right)
    \label{eq:ch4co2gasratio}\\
    \frac{p_{\text{CO}}}{p_{\text{CO}_2}} = \left(\frac{{K}_{\text{H,CO}_2}}{{K}_{\text{H,CO}}}\right)\left(\frac{m_{\text{CO}}}{m_{\text{CO}_2}}\right)
    \label{eq:coco2gasratio}
\end{gather}

To determine the temperature and pressure conditions where gas-liquid partitioning occurs, we began by locating the saturation boundary of water using the Reference Fluid Thermodynamic and Transport Properties (REFPROP) program \citep{Huber_2022}. Then, we generated adiabatic atmospheric pressure-temperature (P-T) profiles to identify where they intersect the saturation boundary. This process is illustrated in Figure \ref{fig:figure2}a and detailed further in Appendix~\ref{sec:PT-profiles}. The intersection between the profiles and the liquid-vapor saturation boundary indicates the approximate location of the cloud deck.

Once the crossing temperature and pressure are determined for each ocean temperature and pressure, we can use the DEW model to evaluate the corresponding Henry's Law constant under saturated conditions. The partial pressure ratios are then solved using Equations~\ref{eq:ch4co2gasratio} and~\ref{eq:coco2gasratio}. Although the gaseous calculations are performed at saturated conditions, they reflect gaseous ratios corresponding to each ``ocean temperature" at the bottom boundary, where chemical equilibrium is reached.



\begin{figure}[ht]
\centering
\includegraphics[width=1\linewidth]{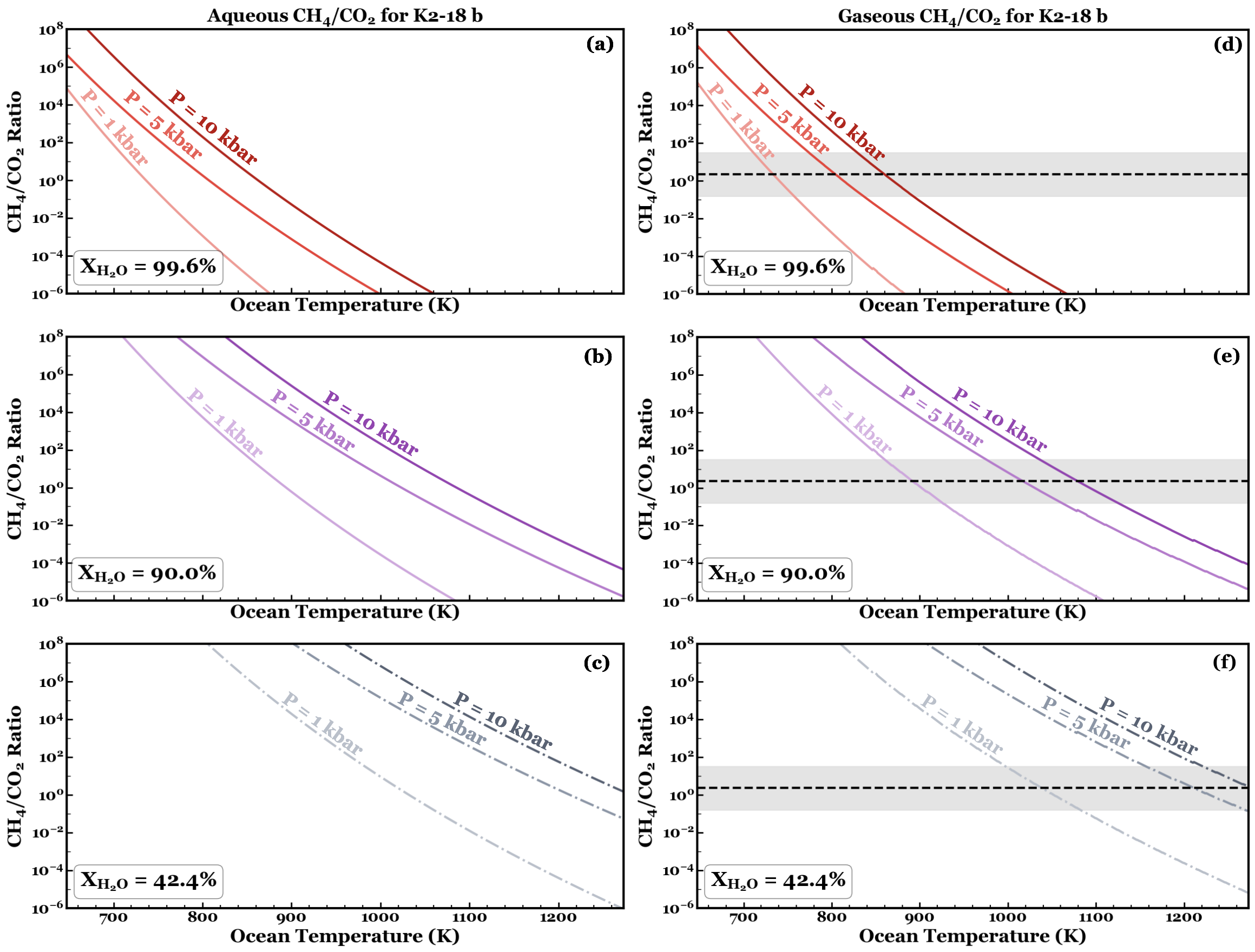}
\caption{Panels (a)-(c) and panels (d)-(f) show the calculated aqueous and the gaseous CH$_4$/CO$_2$ ratios at $X_{\text{H}_2\text{O}}$ = 99.6\%, 90.0\%, and 42.4\%. The $X_{\text{H}_2\text{O}}$ = 42.4\% results are in dashed lines to imply that results are illustrative, as this concentration describes a hydrogen-rich fluid (see Appendix~\ref{sec:assumptions}). The black dashed line and gray region represent the CH$_4$/CO$_2$ ratio and uncertainty presented in \citet{Madhu_Hycean_2023}.
\label{fig:figure3}}
\end{figure}

\section{Results and Discussion} 
\subsection{Can a Supercritical Water Ocean Explain K2-18 b's Atmospheric Chemistry?}
\begin{figure}[ht!]
\centering
\includegraphics[width=1\linewidth]{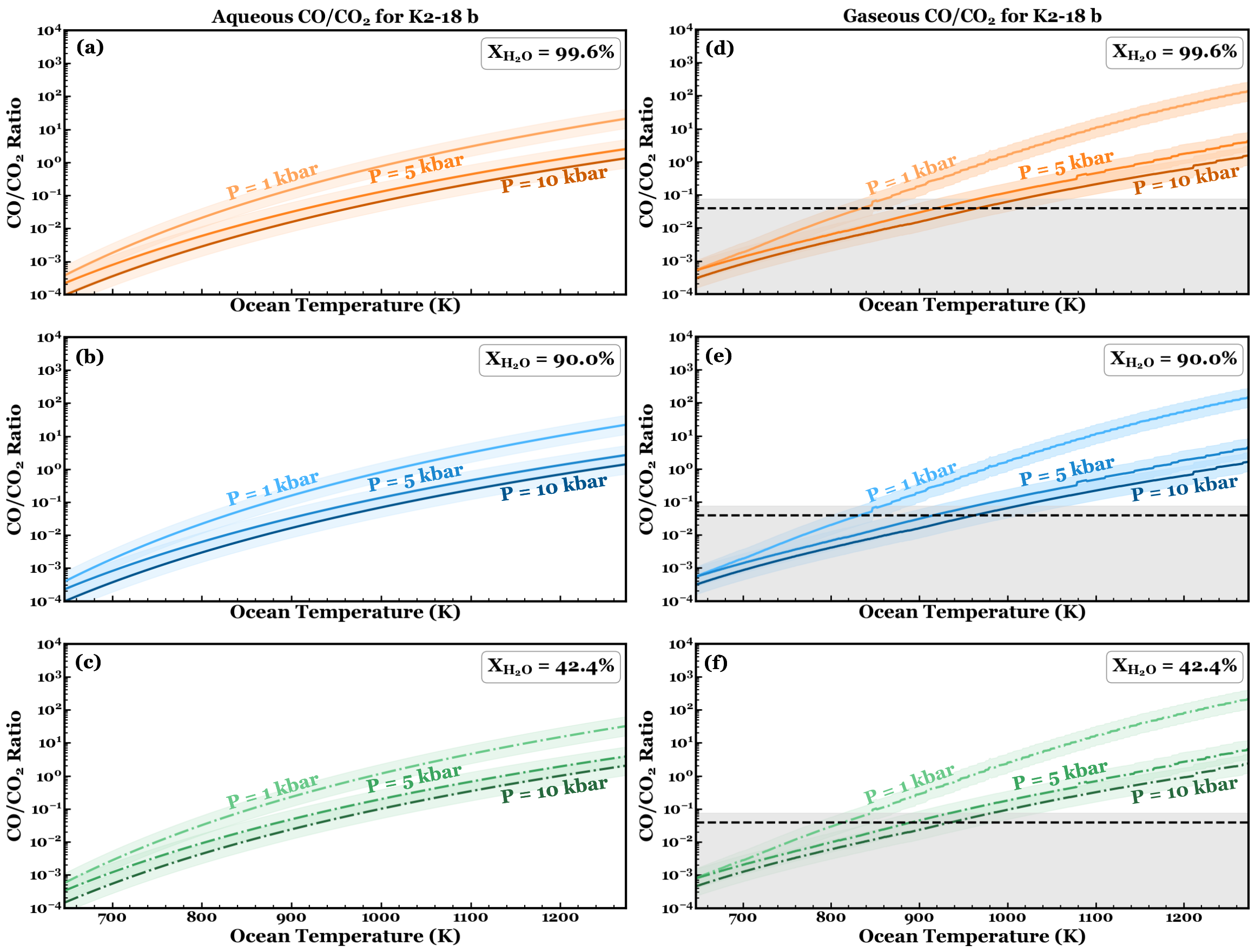}
\caption{(a-c) Aqueous and (d-f)  gaseous CO/CO$_2$ ratios. The symbols and the colors are labeled the same ways as in Figure~\ref{fig:figure3}. The region of uncertainty surrounding the curves is due to the one-offset error for the CH$_4$/CO$_2$ ratio.}
\label{fig:figure4}
\end{figure}

Our modeling approach examines both the aqueous and gas phases to evaluate the impact of phase transitions on the ratios of carbon species in K2-18 b's hypothetical supercritical water ocean and in its atmosphere. However, we focus primarily on the results for the gaseous ratios since they correspond most closely to the observed abundances in K2-18 b's upper atmosphere. For the CH$_4$/CO$_2$ ratio (Figure \ref{fig:figure3}), our results show that higher ocean temperatures favor the production of CO$_2$, while higher pressures and lower water mole fractions (more H$_2$) promote the formation of CH$_4$. We find that the predicted gaseous CH$_4$/CO$_2$ ratios indicate a lower temperature limit of T $\approx$ 710 K to reproduce the observed CH$_4$/CO$_2$ ratio, across the full range of H$_2$ molalities explored. Sub-critical temperatures would only be consistent with the measured CH$_4$/CO$_2$ ratio if the H$_2$ abundance were very low, requiring K2-18 b to have an unrealistically high water/rock mass ratio. For the CO/CO$_2$ ratio (Figure \ref{fig:figure4}), it can be seen that higher temperatures and lower pressures favor the formation of CO. We find that our gaseous CO/CO$_2$ ratios support temperatures less than $\sim$ 1070 K to avoid exceeding the observational detection limit for CO (log(X$_\text{CO}$) $<-3.50$ at a 95\% upper confidence limit, \citep{Madhu_Hycean_2023}). 



Combining these constraints then provides a self-consistent ocean temperature range of T $\approx$ 710 K -- 1070 K, which exceeds the critical temperatures of pure water as well as relevant H$_2$-H$_2$O mixtures \citep{Seward_1981}. These findings suggest that the presence of a supercritical water ocean can explain the observed carbon ratios in K2-18 b's atmosphere. Between pressures of 1 - 5 kbar, H$_2$ is completely miscible with water according to \cite{Gupta_2024}, leading to a homogeneous supercritical H$_2$-H$_2$O ocean that would categorize K2-18 b as a ``Stratified Mini-Neptune" as defined by \citet{benneke2024jwst}. On the other hand, at pressures between 5 - 10 kbar, H$_2$ and H$_2$O may be immiscible at some temperatures, leading to a phase-separated water-rich layer beneath a supercritical H$_2$-rich layer \citep{Gupta_2024}. We discuss more intricate scenarios and their implications further in Appendix~\ref{subsec:immiscibility}. This structure would categorize K2-18 b as a mini-Neptune similar to Uranus and Neptune interior models \citep{Schiebe_2019, Gupta_2024, Amoros_2024}. Note that unlike ``mini-Neptune'' gas-phase models that assume deep H$_2$-dominated atmospheres for K2-18 b \citep{Hu_2021_Photochemistry, Tsai_2021, Yu_2021, Leconte_2024, Yang_2024}, our supercritical ocean chemistry model can explain the low observed abundance of CO on K2-18 b. 

Both the lower and upper temperature limits are conservative estimates. Higher pressures than those we explored are possible, and if equilibrium is attained at higher pressures, this may shift the lower limit on temperature upwards. In addition, the CO-CO$_2$ couple re-equilibrates faster than CH$_4$-CO$_2$ since the latter reaction requires more bond-breaking and bond-making steps \citep{WEI_2004}, and thus CO may partially re-equilibrate in the warm atmosphere above the ocean. 



We note that the gaseous results show only subtle differences compared to the aqueous phase, for both the CH$_4$/CO$_2$ and CO/CO$_2$ ratios, which align with our expectations, given the relatively similar solubilities of CH$_4$, CO$_2$, and CO in supercritical water. CO$_2$ is slightly more soluble in liquid water than CH$_4$ and CO are, which may cause some fractionation during water cloud condensation. However, this effect diminishes with increasing temperature. As the temperature rises, their respective Henry's Law constants become more similar (see Figure~\ref{fig:appendixfig1}). This reduces their differences in solubility and minimizes fractionation. As a result, the ratios in the gaseous phase closely mirror those of the aqueous phase, with only minor variations driven by the slightly higher solubility of CO$_2$.   

Although our gaseous calculations represent carbon ratios at the cloud interface layer, we can directly compare these results with observed abundances from the region probed by JWST. We assume the gaseous ratios remain largely unchanged for two main reasons. First, previous photochemical models suggest that these carbon species are quenched at pressures between 10-100 bar \citep{Hu_2021_Photochemistry, Tsai_2021, Yu_2021, Yang_2024}, similar to the pressure levels at the base of the cloud layer where our calculations are performed. At pressures lower than the quench point, the mixing ratios of these species remain constant as they are transported upward through the atmosphere until photochemistry becomes important. Since JWST primarily probes the atmosphere at pressures between 0.1-2 mbar \citep{Rustamkulov_2023} and multiple photochemical models demonstrate that photochemistry may have minimal influence on the atmospheric abundances within this region \citep{Tsai_2021, Yu_2021, Tsai_2024, Huang_2024, Yang_2024}, our calculations are well-suited for comparison with JWST data. Secondly, the relative insolubility of these carbon species in liquid water clouds suggests that their gaseous ratios should undergo minimal change as they move through the water cloud. As a result, our calculated gaseous ratios at the interface with the cloud layer can be directly compared with JWST observations.


\subsection{Implications for Habitability}

A key question surrounding K2-18 b is whether it can host a habitable environment. Atmospheric observations suggest the ocean temperature should be at least 710 K to explain the observed carbon ratios, assuming thermochemical equilibrium controls dissolved gas abundances in a supercritical water ocean. Currently, the theoretical upper-temperature limit for microbial growth is considered to be $\sim$ 150$^{\circ}$C (423.15 K) \citep{Schulze-Makuch_2017}, indicating that a supercritical water ocean cannot support life as we know it. However, life existing solely within the clouds is a possibility and has been explored for sub-Neptunes by \citet{Seager_2021}.  Although a supercritical water ocean may not permit habitability, studying its chemistry offers a unique opportunity to understand exotic environments on exoplanets like K2-18 b. 

A global supercritical water ocean could foster a hydrothermal environment characterized by hot aqueous fluids. Supercritical fluids have been observed in hydrothermal vent systems on Earth \citep{VONDAMM_2003, Koschinsky_2008}, however, we are not implying the presence of individual hydrothermal vents on a rocky seafloor on K2-18 b. Instead, K2-18 b may be a planetary-scale hydrothermal system unlike anything known on Earth.  Beyond our planet, similar hydrothermal activity has only been implicated by data from Saturn's moon Enceladus \citep{Hsu_2015, Waite_2017}, and hypothesized on Jupiter's moon Europa \citep{Daswani_2021, behounkova_2024}, with no evidence found on an exoplanet to date. While our study does not confirm the existence of hydrothermal processes on K2-18 b, it motivates further investigation into this possibility. The methodology developed here can serve as a first step to investigating hydrothermal worlds.


\subsection{Future Work}
Our finding is consistent with previous climate models of sub-Neptunes, which suggest that water at the lower boundary would likely exist in a supercritical state rather than a liquid state \citep{Innes_2023, Pierrehumbert_2023, Leconte_2024}. Having outlined constraints for a hypothetical supercritical water ocean on K2-18 b, we now turn to the next steps toward an improved understanding of its interior using atmospheric observations.

The most well-constrained abundances from current JWST observations are those of CH$_4$ and CO$_2$, which yield a CH$_4$/CO$_2$ ratio on the order of unity \citep{Madhu_Hycean_2023}. As these measurements are refined with new data, shifts in this ratio within the error range could occur but may still support the possibility of a supercritical water ocean. Current atmospheric models consistent with the CH$_4$ and CO$_2$ observations include a thin-atmosphere liquid water world with biological CH$_4$ flux, a mini-Neptune lacking a surface influence, and the supercritical hydrothermal water world modeled here. To distinguish between these possibilities, the abundances of CO and NH$_3$ then become critical factors, as each model predicts different levels of these species \citep{Shorttle_2024, wogan2024, Yang_2024}. Our supercritical water ocean model spans a wide range of CO/CO$_2$ ratios, but NH$_3$ was not considered in this study due to the additional complexities involved in modeling its geochemistry. Factors such as ocean pH, total nitrogen inventory, and NH$_3$ dissolution in water clouds introduce greater uncertainties that make NH$_3$ more challenging to model compared to the carbon species. Continued study of the phase behavior of H$_2$-H$_2$O mixtures under conditions relevant to sub-Neptunes and the development of geochemically practical parameterizations would also be useful next steps.


While current models have brought us closer to understanding K2-18 b's atmospheric structure, more precise constraints on CO and NH$_3$ abundances will be instrumental for additional testing and refining of our models. Ongoing and future JWST observations (JWST C1 GO 2372; PI: Renyu Hu, JWST C1 GO 2722; PI: Nikku Madhusudhan) will be crucial in providing these constraints, helping to distinguish between the candidate atmosphere-interior structures.

\section{Conclusion}

Volatile-rich sub-Neptunes have sparked intrigue, with their observed masses and radii hinting at a range of potential interior structures, each carrying distinct implications for the planet's inner workings and the origin of these planets. Recently, transmission spectroscopy of K2-18 b revealed the first detection of CH$_4$ and CO$_2$ in the atmosphere of a sub-Neptune exoplanet, along with a nondetection of CO. These data allow us to evaluate the consistency of various structural models against the observed atmospheric composition. In this work, we explored the possibility of K2-18 b being a global hydrothermal world and found that:


\begin{itemize}
    \item A supercritical water ocean with temperatures ranging from approximately 710 K to 1070 K and pressures between 1 and 10 kbar can explain the CH$_4$/CO$_2$ ratio observed by JWST and the nondetection of CO.
    \item This ocean structure is unlikely to support life as we know it, as the temperature range is far hotter than the upper limit where life has been observed to survive on Earth.
    \item The presence of a supercritical water ocean implies that K2-18 b may host a global hydrothermal environment.
\end{itemize}


Overall, this work corroborates the possibility of K2-18 b being a hydrothermal world and explores how Earth-based geochemical modeling can be a valuable tool for characterizing exoplanet interiors. Future observations that better constrain the abundances of CO and NH$_3$ will be critical for advancing our understanding of K2-18 b and other temperate sub-Neptunes.

\section{Acknowledgements}
X. Yu, C. N. Luu, J. I. Moses, and J. Krissansen-Totton are supported by the NASA Habitable Worlds Program Grant 80NSSC24K0075. C. R. Glein and N. Truong acknowledge support from the Heising-Simons Foundation (grant 2023-4657). X. Yu is supported by the NASA Planetary Science Early Career Award 80NSSC23K1108 and by the Heising-Simons Foundation grant 2023-3936. S.-M.T. is supported by NASA through Exobiology Grant No. 80NSSC20K1437 and Interdisciplinary Consortia for Astrobiology Research (ICAR) Grant Nos. 80NSSC23K1399, 80NSSC21K0905, and 80NSSC23K1398. X. Zhang is supported by the National Science Foundation grant AST2307463 and the NASA Exoplanet Research grant 80NSSC22K0236. A. Aguichine is supported by NASA's Interdisciplinary Consortia for Astrobiology Research (ICAR) Grant No. 80NSSC21K0597. H.I. is supported by the European Union (ERC, DIVERSE, 101087755). This work also benefited from the Exoplanet Summer Programs in the Other Worlds Laboratory (OWL) at the University of California, Santa Cruz, a program funded by the Heising-Simons Foundation. We also acknowledge Dr. Ligia Fonseca Coelho for insightful discussions on the temperature range for microbial growth.

\appendix

\section{Observational data selection}
In this paper, we compare our model predictions against abundances computed using the retrieved one-offset mixing ratio values listed in Table 2 of \citet{Madhu_Hycean_2023}. The offset solutions are preferred as numerous studies have suggested the existence of an offset between the NRS1 and NRS2 detectors of the Near-Infrared Spectrograph (NIRSpec) instrument used to observe K2-18 b \citep{May_2023, Moran_2023}. We propagate the uncertainty of the observed ratios using the largest mixing ratio error given for each species. Given the uncertainties in the atmospheric retrieval, it is desirable to adopt the most conservative posture.

\section{Conversion Between Mass Fraction and Hydrogen Molality}
To convert the mass fraction percentages into hydrogen molalities in Section~\ref{subsec:molality}, we refer to protosolar metallicity values for the H-He mixture consisting of $X_0=70.04$\% H and $Y_0=27.80$\% He \citep{Truong_2024}. Here, $X_0$ and $Y_0$ refer to protosolar mass fractions. If purely H-He are accreted from the protostellar nebula, while excluding all silicate dust and volatile ices, the composition of the gas is 71.58\% H and 28.41\% He by weight. For a planet with a bulk composition determined by the H-He, H$_2$O and rocky core mass fractions $f_\mathrm{H-He}$, $f_\mathrm{H_2O}$ and $f_\mathrm{Rock} = 1 - f_\mathrm{H-He} - f_\mathrm{H_2O}$, respectively, the molality can be computed as follows:

\begin{equation}
    m_{\text{H}_2} = \frac{f_\mathrm{H-He}}{f_\mathrm{H_2O}} \frac{X_0}{X_0+Y_0} \frac{1}{\mu_\mathrm{H_2}},
\end{equation}
where $\mu_\mathrm{H_2}$ stands for the molar mass of H$_2$ in kg$\cdot$(mol$^{-1}$).

\section{Detailed Constraints on Hydrogen Molality}\label{sec:H2_molality}

Here, we explain how the hydrogen molality ($m_{\text{H}_2}$) range used in our modeling was derived. To determine a reasonable range of $m_{\text{H}_2}$ values for our modeling, we first turn toward planet interior structure models to provide a range of possible compositions that match the mass and radius of K2-18 b. This approach provides an independent constraint on the masses of H-He and water, which can be used to estimate the equivalent molality of H$_2$ for the bulk planet. We assume that this value applies to a supercritical water ocean. 

The radius of K2-18 b can be reproduced by end-member cases where the interior is either i) an Earth-like core with a pure H-He envelope and no water, which corresponds to a hydrogen molality ($m_{\text{H}_2}$) of infinity or ii) a nearly 100\% water body with no core and negligible amounts of H-He, which leads to $m_{\text{H}_2} \approx 0$ \citep{Madhusudhan2020}.

A lower limit on the H$_2$ molality can be inferred from the ice-to-rock ratio accreted by the planet. If the composition of the interior is inherited from the protoplanetary disk (PPD), the latter end member described above is a highly unlikely scenario. The ice-to-rock ratio in a planet would be controlled by the composition of pebbles/planetesimals in the disk, which is typically made of a mixture of 50\% ice and 50\% rock (iron and silicates) by mass \citep{Drazkowska2017, Bitsch2020, Schneider2021a, Aguichine2022, Aguichine2024}. Therefore, we place an upper limit on the ice-to-rock content in K2-18 b as 1:1 by mass, similar to the inferred compositions of the icy moons of the Solar System \citep{Sotin2007}. With this upper limit on the bulk water content, K2-18 b requires a minimal amount ($0.02\%$) of H-He to match its radius (see right panel of Figure 3 from \citet{Madhusudhan2020}). Such a composition would correspond to a lower limit $m_{\text{H}_2}=0.14$~mol$\cdot$(kg H$_2$O)$^{-1}$.

Furthermore, sub-Neptunes should have a limit to how much hydrogen they can accrete during formation. The study of \citet{Ginzburg2016} investigated the maximum hydrogen mass fraction retained by the planet according to the core accretion formalism and after an early stage of boil-off. Using their Equation 24, we estimate that the maximum H-He mass fraction accreted by K2-18 b is $\sim 2\%$. However, applying this limit to the interior modeling from \citet{Madhusudhan2020} may or may not constitute an additional constraint depending on the assumptions of the interior model, namely the intrinsic temperature of the planet $T_{\mathrm{int}}$. For $T_{\mathrm{int}}=25$ K, an upper limit of 2\% on the H-He will lead to a bulk composition of 2\% H-He and 18\% H$_2$O. However, for $T_{\mathrm{int}}=50$ K, only 1.5\% of H-He is required to match the bulk density of K2-18 b, meaning that a pure H-He atmosphere cannot be excluded with arguments from interior and formation modeling. Nevertheless, pure H-He envelopes are considered unrealistic, as the hydrogen is expected to react with the FeO component of the silicate mantle to produce H$_2$O \citep{Kite2020, Schlichting2022} and early accretion of planetesimals contribute to the atmosphere's inventory of metals \citep{Fortney_2013, Mol_Lous_2024}. A bulk composition of 2\% H-He and 18\% H$_2$O leads to $m_{\text{H}_2}=42$~mol$\cdot$(kg H$_2$O)$^{-1}$. In addition, \citet{Yang_2024} derived a bulk envelope H$_2$O/H$_2$ ratio through gas-phase chemistry, finding that a scenario in which K2-18 b has 10$ \times Z_{\odot}$ with H$_2$O/H$_2$ = 50/50 is most consistent with observations. This ratio translates to a hydrogen molality of approximately 56.55 mol$\cdot$(kg H$_2$O)$^{-1}$. Since this value is close to our upper limit of 42 mol$\cdot$(kg H$_2$O)$^{-1}$, we choose to adopt 56.55 mol$\cdot$(kg H$_2$O)$^{-1}$ as an upper limit instead. 


\section{Dilute Solution Assumptions in the DEW model}\label{sec:assumptions}
\setcounter{equation}{0} 

Our treatment of supercritical water as a dilute solution involves several assumptions that are implicit in our choice of standard states. A standard state is a set of conditions for a species with a fixed chemical composition, serving as a reference state for calculating thermodynamic properties. Often, the standard state is a hypothetical state that may not be physically possible but remains useful in extrapolating thermodynamic properties. We can relate the activities, molalities, and mole fractions through the definitions of standard states from the following equations, adopted from \citet{Sverjensky_2019}.

For aqueous species, the relationship between the thermodynamic activity and molality is given by,

\begin{gather}
      {a}_{j} = \frac{{\gamma}_{j}{m}_{j}}
      {{\gamma}^{0}_{j}{m}^{0}_{j}}
      \label{eq:aqueous_assumption1}
\end{gather}

where ${\gamma}_{j}$ and ${m}_{j}$ refer to the activity coefficient and molality of species $j$ and ${\gamma}^{0}_{j}$ and ${m}^{0}_{j}$ refer to the standard state values \citep{Anderson_2005}.  We adopt a Henryan standard state (i.e., following Henry's Law) known as the ``ideal one molal state" for aqueous species. This is a commonly used hypothetical standard state that describes an ideal solution at infinite dilution with ${m}^{0}_{j}$ = 1.0. By definition, real aqueous species cannot physically exist in this state.  Under this condition, since ${m}^{0}_{j}$ = 1.0 and ${\gamma}^{0}_{j}$ = 1.0 (see Figure 5 of \citet{Sverjensky_2019}), Equation~\ref{eq:aqueous_assumption1} becomes,


\begin{gather}
      {a}_{j} = \frac{{\gamma}_{j}{m}_{j}}
      {(1.0)(1.0)} = {\gamma}_{j}{m}_{j}
    \label{eq:aqueous_assumption2}
\end{gather}

The ratio of activities for species $j$ and $i$ becomes,

\begin{gather}
      \frac{{a}_{j}}{{a}_{i}} = \frac{{\gamma}_{j}{m}_{j}}
      {{\gamma}_{i}{m}_{i}}
      \label{eq:aqueous_assumption3}
\end{gather}

The species of interest in this paper (CH$_4$, CO$_2$, and CO) are neutral, nonpolar molecules, and while their activity coefficients may depart from unity, the ratio of their activity coefficients tends to cancel out (e.g., \citealt{BARRETT_1988}). As a result, the ratios of their activities can be approximated as,

\begin{gather}
      \frac{{a}_{j}}{{a}_{i}} \approx \frac{{m}_{j}}
      {{m}_{i}}
      \label{eq:aqueous_assumption4}
\end{gather}

For water, we adopt a Raoultian standard state (i.e., following Raoult's Law) where the fluid phase is pure at some temperature and pressure, and the activity of the ideal solution is equal to its mole fraction. This is the simplest standard state, representing the actual pure phase of the species \citep{Anderson_2005}. Here, the activity of water is related to the mole fraction by the equation,

\begin{gather}
    {a}_{\text{H}_2\text{O}} = {\lambda}_{\text{H}_2\text{O}}{X}_{\text{H}_2\text{O}}
    \label{eq:water_assumption1}
\end{gather}

where ${\lambda}_{\text{H}_2\text{O}}$ represents a rational activity coefficient.  In the present system, water behaves ideally and ${\lambda}_{\text{H}_2\text{O}}$ $\rightarrow$ 1.0 as ${X}_{\text{H}_2\text{O}}$ $\rightarrow$ 1.0, and therefore,

\begin{gather}
    {a}_{\text{H}_2\text{O}} \approx {X}_{\text{H}_2\text{O}}
    \label{eq:water_assumption2}
\end{gather}

The results of Equations~\ref{eq:aqueous_assumption4} and ~\ref{eq:water_assumption2} are hence substituted into Equations \ref{eq:equilibrium1} and \ref{eq:equilibrium2}.

\section{Generating the Pressure-Temperature Profiles}\label{sec:PT-profiles}

To generate the P-T profiles, we assume that carbon species equilibrate in a supercritical water ocean at some depth and are then transported upward by turbulent mixing. Eventually, on K2-18 b, these species encounter liquid water droplets higher in the atmosphere. Dissolution into the droplets can change the ratios of species that remain in the gas phase. The temperature profiles at high pressure were assumed to lie on a pure water adiabat. Values for the adiabatic temperature gradient were calculated using the AQUA equation of state for water \citep{haldemann2020}, which provides this quantity tabulated as a function of temperature and pressure. Different profiles were generated by integrating upwards from different starting points in the supercritical region of water's phase diagram. The range of profiles reflects the uncertainty in the equilibrium temperature at depth, which would require a more comprehensive coupled radiative transfer and interior model to be determined accurately.

Care was taken in proximity to the liquid-vapor phase boundary, where a discontinuous jump in the adiabatic temperature gradient can produce errors in the data interpolation. Integration was stopped when the temperature was within 10 K of the phase boundary, and the profiles were then linearly interpolated (in logarithmic pressure space) towards the phase boundary. Warmer adiabats that do not intersect this boundary will have hydrogen in much lower concentrations in the upper atmosphere since in this case there is no cold trapping of water and the hydrogen will be fully miscible with supercritical water \citep{soubiran2015}, which is not consistent with the non-detection of water in K2-18 b's atmosphere \citep{Madhu_Hycean_2023}.


\counterwithin{figure}{section}
\section{Parameters for Equilibrium Calculations}

The equilibrium constants and Henry's Law constants in Equations \ref{eq:equilibrium1}, \ref{eq:equilibrium2}, \ref{eq:henrys1}, \ref{eq:henrys2}, and \ref{eq:henrys3} are obtained with the DEW model, which uses a series of revised Helgeson-Kirkham-Flowers (HKF) equations of state for aqueous species \citep{Shock_1988, SHOCK_HELGESON_1990, SVERJENSKY2014125} and an equation of state by \cite{ZHANG_2005} for water. For convenience, the computed values are then fitted with the logarithmic equation and associated values in Table \ref{tab:chemicals}. The equilibrium constants in Equations \ref{eq:equilibrium1} and \ref{eq:equilibrium2} are fitted on a linear grid of temperatures between 647.15 K - 1273.15 K at every 1 K following the lower limit, resembling the ocean temperature, at each surface pressure (1000 bar, 5000 bar, 10,000 bar). 

\begin{figure}[ht]
\centering
\includegraphics[width=0.7\linewidth]{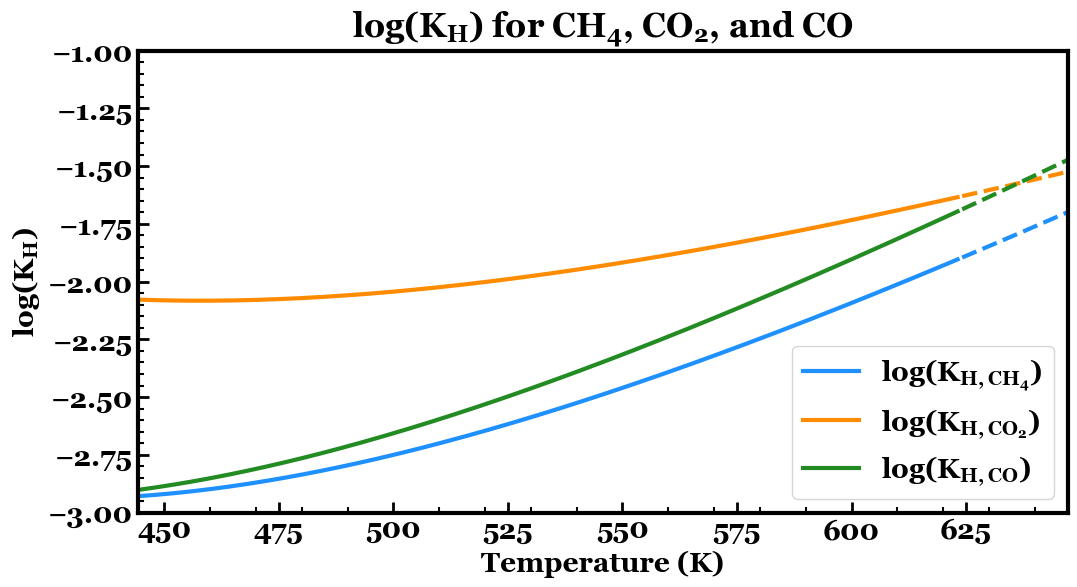}
\caption{Logarithmicly fitted Henry's Law constants for CH$_4$, CO$_2$, and CO across temperatures between 444.15 K to 623.15 K represented by solid lines, with extrapolated values between 624.15 K to 647.15 K represented by dashed lines. Pressures correspond to water-saturated conditions as shown in Figure~\ref{fig:figure2}a.
\label{fig:appendixfig1}}
\end{figure}

The Henry's Law constants are initially fitted on a linear grid of temperatures between 444.15 K to 623.15 K at every 1 K. The lower limit represents the lowest temperature at which our P-T profiles cross the saturation boundary, while the upper limit represents DEW's calculation limit at 350$^{\circ}$C (623.15 K). To account for our P-T profiles with saturation temperatures reaching up to 647.15 K, we extrapolate the Henry's Law constants using the logarithmic equation shown in Table \ref{tab:chemicals} based on the fitting between 444.15 K - 623.15 K. Here, the corresponding pressures are the water saturated vapor pressures, referred to as ``P$_\text{sat}$ Conditions". Although this extrapolation introduces some uncertainty, the fitting achieves R$^2$ values of at least 0.996, demonstrating a close fit to the data. Figure~\ref{fig:appendixfig1} shows the trend of Henry's Law constants across the range of temperatures.

\setcounter{table}{0}
\renewcommand{\thetable}{F\arabic{table}}

\begin{deluxetable*}{|c|c|c|c|CCC|}
\tabletypesize{\scriptsize}
\tablewidth{0pt}
\tablecaption{Summary of Key Chemical Reactions and Parameters to Calculate Equilibrium Constants
\label{tab:chemicals}}
\tablehead{
\colhead{Name} & 
\colhead{Reaction} & 
\colhead{Pressure (kbar)} &
\colhead{$R^2$} & 
\multicolumn{3}{c}{$\log(K) = A + \frac{B}{T} + C\log(T)$} \\
\cline{5-7}
\colhead{} &
\colhead{} &
\colhead{} &
\colhead{} &
\colhead{A} & 
\colhead{B} & 
\colhead{C}
}
\startdata
\multirow{3}{*}{Methane-Carbon Dioxide} & \multirow{3}{*}{CO$_2$(aq) + 4H$_2$(aq) $\rightarrow$ CH$_4$(aq) + 2H$_2$O(sc)} & 1 & 0.998 & 81.668 & 15682.665 & -34.718 \\
\cline{3-7}
& & 5 & 0.999 & 127.989 & 7364.805 & -45.991 \\
\cline{3-7} 
& & 10 & 0.999 & 120.328 & 9699.616 & -43.654 \\
\hline 
\multirow{3}{*}{Carbon Monoxide-Carbon Dioxide} & \multirow{3}{*}{CH$_4$(aq) + 4CO$_2$(aq) $\rightarrow$ CO$_2$(aq) + 4CO(aq) + 2H$_2$O(sc)} & 1 & 0.999 & -36.958 & -18001.432 & 18.046 \\
\cline{3-7}
& & 5 & 0.999 & -72.516 & -11110.618 & 26.572 \\
\cline{3-7} 
& & 10 & 0.999 & -67.254 & -12232.227 & 24.800 \\
\hline 
Henry's Law for Methane & CH$_4$(g) $\rightarrow$ CH$_4$(aq) & P$_\text{sat}$ Conditions & 0.997 & -120.131 & 7106.969 & 38.224 \\
\hline 
Henry's Law for Carbon Dioxide & CO$_2$(g) $\rightarrow$ CO$_2$(aq) & P$_\text{sat}$ Conditions & 0.996 & -76.414 & 4775.558 & 24.016 \\
\hline
Henry's Law for Carbon Monoxide & CO(g) $\rightarrow$ CO(aq) & P$_\text{sat}$ Conditions & 0.998 & -121.555 & 6977.213 & 38.883 \\
\enddata
\tablecomments{This table presents the parameters for various chemical reactions. The constants A, B, and C are used in the logarithmic equation $\log(K) = A + \frac{B}{T} + C\log(T)$, where T refers to the absolute temperature. }
\end{deluxetable*}

\section{Caveats}
\label{subsec:caveats}
\subsection{P-T Profiles of Hydrogen-Water Mixtures}
To address the challenges of modeling K2-18 b's interior, we employed a geochemical model to investigate the chemistry of a hypothesized supercritical water ocean on K2-18 b. However, the existing DEW model is limited to water-rich fluids (see the assumptions in Appendix~\ref{sec:assumptions}) and cannot fully account for scenarios with higher hydrogen molalities. It is possible that K2-18 b possesses significantly higher hydrogen molalities than those modeled here. An upper limit on $m_{\text{H}_2}$ can be inferred from JWST-derived atmospheric abundances reported by \citet{Madhu_Hycean_2023}. The volume mixing ratios of CH$_4$ and CO$_2$, each at approximately 1\%, provide a lower bound on the bulk volatile content, thus constraining the upper limit on the hydrogen content. Using water as a proxy for all volatiles, this corresponds to an H$_2$ molality upper limit of 1300 mol$\cdot$(kg H$_2$O)$^{-1}$. Such a high H$_2$ molality would represent an extremely water-poor fluid, making it incompatible with the present model. To investigate these possible H$_2$-rich scenarios, conventional gas-phase models \citep{Yang_2024} and/or molecular dynamics (MD) simulations such as those used in \citet{Gupta_2024} and \citet{Amoros_2024} may be more appropriate. 

This potential limitation also affects our P-T profiles and the saturation boundary in Figure~\ref{fig:figure2}a, and consequently their intersections, where we performed Henry's Law calculations. In this work, both the P-T profiles and the saturation boundary were constructed for pure water, or $X_{\text{H}_2\text{O}}$ = 100\%. However, K2-18 b's lower atmosphere has $X_{\text{H}_2\text{O}}$ $<$ 100\%. In addition, it should be noted that we reference the critical point of pure water rather than that of an H$_2$-H$_2$O mixture. To estimate how different H$_2$-H$_2$O mixtures might influence the saturation boundary -- and, by extension, the crossing temperatures and pressures -- we analyzed the saturation boundary corresponding to each mole fraction of water used in this work. We used the Reference Fluid Thermodynamic and Transport Properties (REFPROP) program \citep{LEMMON-RP10} to compare the difference in saturation temperatures up to the critical temperature of pure water. The results are summarized in Table~\ref{tab:comparison}. We find that these temperature changes are not significant enough to greatly impact our results, given other uncertainties such as the fluid oxidation state (see Appendix~\ref{sec:assumptions}). We note that this calculation does not include non-ideal gas behavior which becomes more prominent at higher pressures and consequently shifts the crossing temperature, however, changes in crossing temperature also do not matter much in the C-O-H system because solubility corrections applied to exsolved gases are similar (Figure~\ref{fig:appendixfig1}). Nevertheless, future work should consider the impact of non-ideal phases and incorporate P-T profiles and saturation boundaries that account for the effects of species other than water to achieve a more self-consistent model.


\subsection{Carbon Dioxide Mixing Ratio}
We assume that gaseous calculations performed for the liquid-vapor interface (cloud bottom) can be directly compared to the JWST observations. However, one notable exception may be CO$_2$, which can remain in a kinetically controlled balance with quenched CO and H$_2$O at altitudes above the CO-CH$_4$-H$_2$O quench point. In that situation, the CO$_2$ mixing ratio can often increase with altitude above the quench point, resulting in a CO$_2$ mole fraction at higher altitudes in excess of expectations based on its deeper equilibrium abundance \citep{Yu_2021,Hu_2021_Photochemistry}. Further disequilibrium chemistry gas-phase modeling may be needed to test this scenario.

\subsection{Hydrogen-Water Immiscibility?}
\label{subsec:immiscibility}
The pressures and temperatures we explored in this work partially overlap with a region of possible H$_2$-H$_2$O immiscibility between 5 kbar and 10 kbar, as shown in Figure 3 of \citet{Gupta_2024}. Instead of a homogeneous layer, H$_2$-H$_2$O immiscibility would lead to the formation of two phase-separated layers: an H$_2$-rich layer on top of an H$_2$O-rich layer. We performed additional calculations investigating how such phase separation could change our ocean temperature lower limit, which is currently given by the CH$_4$/CO$_2$ ratio. In this case, the equilibrium CH$_4$/CO$_2$ ratio is determined by the H$_2$ molality and mole fraction of water in the H$_2$O-rich layer on the bottom. If H$_2$ is removed, this layer is expected to contain less dissolved hydrogen than a miscible H$_2$/H$_2$O layer, leading to a higher water mole fraction. Thus, we performed sensitivity tests self-consistently varying the mole fraction of water in our model between our highest water mole fraction, $X_{\text{H}_2\text{O}}$ = 99.6\%, up to but not including $X_{\text{H}_2\text{O}}$ = 100.0\% for our CH$_4$/CO$_2$ ratios at 5 kbar, the lowest pressure for possible immiscibility within our temperature range \citep{Gupta_2024}. We found that our modeled CH$_4$/CO$_2$ ratios still agree with the atmospheric observations, given an ocean temperature lower limit of 710 K, when the mole fraction of water is less than $X_{\text{H}_2\text{O}}$ = 99.94\% ($X_{\text{H}_2}$ $\approx$ 0.06\% or $m_{\text{H}_2}$ $\approx$ 0.03 mol$\cdot$(kg H$_2$O)$^{-1}$). However, the lower temperature limit is likely to be immune to the effects of phase splitting, as it is determined by the most water-rich composition (Figure 3a). Miscibility can be expected at an H$_2$ molality of 0.14 mol$\cdot$(kg H$_2$O)$^{-1}$. To decrease our current estimated lower ocean temperature limit below $\sim$ 710 K, the H$_2$ molality would need to decrease by a factor of $\sim$5 in a phase-separated H$_2$O-rich layer to achieve $X_{\text{H}_2\text{O}}$ = 99.94\%. Meanwhile, to investigate how our upper limit on temperature may change due to higher water mole fractions in the phase-separated H$_2$O-rich layer, we performed similar sensitivity tests for our CO/CO$_2$ ratios at 10 kbar, the highest pressure of interest. Our CO/CO$_2$ ratios are minimally changed since, as the water mole fraction approaches unity, the proportional increase in CO/CO$_2$ reaches diminishing returns (Equation~\ref{eq:equilibrium2}).

\subsection{Supercritical-Liquid Water Adiabats}
We find that a portion of our P-T profiles go directly from the supercritical phase to the liquid water phase, as seen by the leftmost adiabat for the 1 kbar pressure in Figure~\ref{fig:figure2}a. This implies a potential interior structure consisting of a hot liquid water layer above a supercritical water layer. However, it is important to note that our P-T profiles do not fully include climate considerations. As mentioned in Section 1, past climate models provide a strong case that surface water on K2-18 b would exist in the supercritical state. \citet{Innes_2023} considered sub-Neptunes with cloud-free, pure H/He atmospheres receiving solar flux from M stars, and found that an atmospheric pressure of 10 bars will lead to a runaway greenhouse effect due to collision-induced absorption. As a result of the runaway greenhouse effect, any liquid water ocean would become supercritical. It has been suggested that surface water can remain liquid in the presence of clouds that provide the necessary albedo \citep{Madhu_Hycean_2023}. Recent studies argue that the high albedo required ($>$ 0.5) is both difficult to achieve for K2-18's spectral type and inconsistent with observational data \citep{Leconte_2024}, although such a high albedo cannot be entirely ruled out (see \citet{Cooke_2024}). These findings support previous interior models suggesting that water-rich sub-Neptunes with a runaway greenhouse effect will develop thick gaseous envelopes with supercritical water \citep{Mousis_2020}. Therefore, when considering these climate models, we believe it is unlikely for surface water to remain in a liquid state. Our conclusion regarding habitability remains the same for either structure since all adiabats directly entering the liquid phase persist at temperatures above 500 K.

\subsection{Species Abundances}
Our approach allows us to discern the CH$_4$/CO$_2$ and CO/CO$_2$ ratios that are consistent with observations. However, it is not yet applicable for calculating the mixing ratios of each species as this requires the detection of carbon-bearing molecules with a higher number of carbon atoms. If future observations detect molecules with more carbon atoms, we can begin to predict the absolute abundances with the DEW model.

\setcounter{table}{0}
\renewcommand{\thetable}{G\arabic{table}}

\begin{deluxetable}{ccccc}
\tablecaption{Comparison of H$_2$-H$_2$O Compositions and Critical Properties\label{tab:comparison}}
\tablewidth{1\textwidth} 
\tablehead{
\colhead{$X_{\text{H}_2\text{O}}$} & 
\colhead{$X_{\text{H}_2}$} & 
\colhead{Maximum Difference in T$_\text{sat}$ (K)} &
\colhead{$X_{\text{H}_2}$} & 
\colhead{Critical Temperature (K) }
 \\
\colhead{(This work)} & 
\colhead{(This work)} & 
\colhead{(This work)} & 
\colhead{\citep{Seward_1981}} & 
\colhead{\citep{Seward_1981}} 
}
\startdata
99.6\% & 0.2\% & 0.26 & 0.5\% & 647.45 \\
90.0\% & 10.0\% & 8.79 & 9.0\% & 647.65 \\
42.4\%* & 43.3\%* & 67.80 & 40.0\% & 656.45 \\
\enddata
\tablecomments{*Derived from \cite{Yang_2024} and used in this work. We also note that not all of the H$_2$O and H$_2$ mole fractions used here sum to unity because of the addition of helium, while \citet{Seward_1981} conducted experiments with H$_2$-H$_2$O mixtures excluding helium. The maximum differences from saturated temperatures in this work are computed from the saturation boundary and critical temperature of pure water. Water clouds form at lower temperatures as the atmosphere becomes poorer in water vapor content}
\end{deluxetable}

\newpage
\bibliographystyle{aasjournal}
\bibliography{refs}

\begin{thebibliography}{}
\expandafter\ifx\csname natexlab\endcsname\relax\def\natexlab#1{#1}\fi
\providecommand{\url}[1]{\href{#1}{#1}}
\providecommand{\dodoi}[1]{doi:~\href{http://doi.org/#1}{\nolinkurl{#1}}}
\providecommand{\doeprint}[1]{\href{http://ascl.net/#1}{\nolinkurl{http://ascl.net/#1}}}
\providecommand{\doarXiv}[1]{\href{https://arxiv.org/abs/#1}{\nolinkurl{https://arxiv.org/abs/#1}}}

\bibitem[{Aguichine \& Mousis(2024)}]{Aguichine2024}
Aguichine, A., \& Mousis, O. 2024

\bibitem[{{Aguichine} {et~al.}(2022){Aguichine}, {Mousis}, \&
  {Lunine}}]{Aguichine2022}
{Aguichine}, A., {Mousis}, O., \& {Lunine}, J.~I. 2022, \psj, 3, 141,
  \dodoi{10.3847/PSJ/ac6bf1}

\bibitem[{Anderson(2005)}]{Anderson_2005}
Anderson, G.~M. 2005, Thermodynamics of Natural Systems, 2nd edn. (Cambridge
  University Press)

\bibitem[{Barrett {et~al.}(1988)Barrett, Anderson, \& Lugowski}]{BARRETT_1988}
Barrett, T., Anderson, G., \& Lugowski, J. 1988, Geochimica et Cosmochimica
  Acta, 52, 807, \dodoi{https://doi.org/10.1016/0016-7037(88)90352-3}

\bibitem[{{Batalha}(2014)}]{Batalha_2014}
{Batalha}, N.~M. 2014, Proceedings of the National Academy of Science, 111,
  12647, \dodoi{10.1073/pnas.1304196111}

\bibitem[{B{\v e}hounkov{\'a} {et~al.}(2021)B{\v e}hounkov{\'a}, Tobie,
  Choblet, Kervazo, Melwani~Daswani, Dumoulin, \& Vance}]{behounkova_2024}
B{\v e}hounkov{\'a}, M., Tobie, G., Choblet, G., {et~al.} 2021, Geophysical
  Research Letters, 48, e2020GL090077,
  \dodoi{https://doi.org/10.1029/2020GL090077}

\bibitem[{{Benneke} {et~al.}(2019){Benneke}, {Wong}, {Piaulet}, {Knutson},
  {Lothringer}, {Morley}, {Crossfield}, {Gao}, {Greene}, {Dressing},
  {Dragomir}, {Howard}, {McCullough}, {Kempton}, {Fortney}, \&
  {Fraine}}]{Benneke_2019}
{Benneke}, B., {Wong}, I., {Piaulet}, C., {et~al.} 2019, \apjl, 887, L14,
  \dodoi{10.3847/2041-8213/ab59dc}

\bibitem[{Benneke {et~al.}(2024)Benneke, Roy, Coulombe, Radica, Piaulet, Ahrer,
  Pierrehumbert, Krissansen-Totton, Schlichting, Hu, Yang, Christie, Thorngren,
  Young, Pelletier, Knutson, Miguel, Evans-Soma, Dorn, Gagnebin, Fortney,
  Komacek, MacDonald, Raul, Cloutier, Acuna, Lafrenière, Cadieux, Doyon,
  Welbanks, \& Allart}]{benneke2024jwst}
Benneke, B., Roy, P.-A., Coulombe, L.-P., {et~al.} 2024, JWST Reveals CH$_4$,
  CO$_2$, and H$_2$O in a Metal-rich Miscible Atmosphere on a Two-Earth-Radius
  Exoplanet.
\newblock \doarXiv{2403.03325}

\bibitem[{{Bitsch} \& {Battistini}(2020)}]{Bitsch2020}
{Bitsch}, B., \& {Battistini}, C. 2020, \aap, 633, A10,
  \dodoi{10.1051/0004-6361/201936463}

\bibitem[{{Cano Amoros} {et~al.}(2024){Cano Amoros}, {Nettelmann}, {Tosi},
  {Baumeister}, \& {Rauer}}]{Amoros_2024}
{Cano Amoros}, M., {Nettelmann}, N., {Tosi}, N., {Baumeister}, P., \& {Rauer},
  H. 2024, arXiv e-prints, arXiv:2410.21099, \dodoi{10.48550/arXiv.2410.21099}

\bibitem[{{Cloutier} {et~al.}(2019){Cloutier}, {Astudillo-Defru}, {Doyon},
  {Bonfils}, {Almenara}, {Bouchy}, {Delfosse}, {Forveille}, {Lovis}, {Mayor},
  {Menou}, {Murgas}, {Pepe}, {Santos}, {Udry}, \&
  {W{\"u}nsche}}]{Cloutier_2019}
{Cloutier}, R., {Astudillo-Defru}, N., {Doyon}, R., {et~al.} 2019, \aap, 621,
  A49, \dodoi{10.1051/0004-6361/201833995}

\bibitem[{{Cooke} \& {Madhusudhan}(2024)}]{Cooke_2024}
{Cooke}, G.~J., \& {Madhusudhan}, N. 2024, arXiv e-prints, arXiv:2410.07313,
  \dodoi{10.48550/arXiv.2410.07313}

\bibitem[{{Drazkowska} \& {Alibert}(2017)}]{Drazkowska2017}
{Drazkowska}, J., \& {Alibert}, Y. 2017, \aap, 608, A92,
  \dodoi{10.1051/0004-6361/201731491}

\bibitem[{Fortney {et~al.}(2013)Fortney, Mordasini, Nettelmann, Kempton,
  Greene, \& Zahnle}]{Fortney_2013}
Fortney, J.~J., Mordasini, C., Nettelmann, N., {et~al.} 2013, The Astrophysical
  Journal, 775, 80, \dodoi{10.1088/0004-637X/775/1/80}

\bibitem[{Fressin {et~al.}(2013)Fressin, Torres, Charbonneau, Bryson,
  Christiansen, Dressing, Jenkins, Walkowicz, \& Batalha}]{Fressin_2013}
Fressin, F., Torres, G., Charbonneau, D., {et~al.} 2013, The Astrophysical
  Journal, 766, 81, \dodoi{10.1088/0004-637X/766/2/81}

\bibitem[{{Fulton} {et~al.}(2017){Fulton}, {Petigura}, {Howard}, {Isaacson},
  {Marcy}, {Cargile}, {Hebb}, {Weiss}, {Johnson}, {Morton}, {Sinukoff},
  {Crossfield}, \& {Hirsch}}]{Fulton_2017}
{Fulton}, B.~J., {Petigura}, E.~A., {Howard}, A.~W., {et~al.} 2017, \aj, 154,
  109, \dodoi{10.3847/1538-3881/aa80eb}

\bibitem[{{Ginzburg} {et~al.}(2016){Ginzburg}, {Schlichting}, \&
  {Sari}}]{Ginzburg2016}
{Ginzburg}, S., {Schlichting}, H.~E., \& {Sari}, R. 2016, \apj, 825, 29,
  \dodoi{10.3847/0004-637X/825/1/29}

\bibitem[{Glein(2024)}]{Glein_2024}
Glein, C.~R. 2024, The Astrophysical Journal Letters, 964, L19,
  \dodoi{10.3847/2041-8213/ad3079}

\bibitem[{{Glein} {et~al.}(2008){Glein}, {Zolotov}, \& {Shock}}]{Glein_2008}
{Glein}, C.~R., {Zolotov}, M.~Y., \& {Shock}, E.~L. 2008, \icarus, 197, 157,
  \dodoi{10.1016/j.icarus.2008.03.021}

\bibitem[{{Gupta} {et~al.}(2024){Gupta}, {Stixrude}, \&
  {Schlichting}}]{Gupta_2024}
{Gupta}, A., {Stixrude}, L., \& {Schlichting}, H.~E. 2024, arXiv e-prints,
  arXiv:2407.04685, \dodoi{10.48550/arXiv.2407.04685}

\bibitem[{{Haldemann} {et~al.}(2020){Haldemann}, {Alibert}, {Mordasini}, \&
  {Benz}}]{haldemann2020}
{Haldemann}, J., {Alibert}, Y., {Mordasini}, C., \& {Benz}, W. 2020, Astronomy
  \& Astrophysics, 643, A105, \dodoi{10.1051/0004-6361/202038367}

\bibitem[{Hsu {et~al.}(2015)Hsu, Postberg, Sekine, Shibuya, Kempf, Horányi,
  Juhász, Altobelli, Suzuki, Masaki, \& et~al.}]{Hsu_2015}
Hsu, H.-W., Postberg, F., Sekine, Y., {et~al.} 2015, Nature, 519, 207,
  \dodoi{10.1038/nature14262}

\bibitem[{{Hu}(2021)}]{Hu_2021_Photochemistry}
{Hu}, R. 2021, \apj, 921, 27, \dodoi{10.3847/1538-4357/ac1789}

\bibitem[{Hu {et~al.}(2021)Hu, Damiano, Scheucher, Kite, Seager, \&
  Rauer}]{Hu_2021_Thermochemistry}
Hu, R., Damiano, M., Scheucher, M., {et~al.} 2021, The Astrophysical Journal
  Letters, 921, \dodoi{10.3847/2041-8213/ac1f92}

\bibitem[{Huang {et~al.}(2024)Huang, Yu, Tsai, Moses, Ohno, Krissansen-Totton,
  Zhang, \& Fortney}]{Huang_2024}
Huang, Z., Yu, X., Tsai, S.-M., {et~al.} 2024, The Astrophysical Journal, 975,
  146, \dodoi{10.3847/1538-4357/ad76ac}

\bibitem[{Huber {et~al.}(2022)Huber, Lemmon, Bell, \& McLinden}]{Huber_2022}
Huber, M.~L., Lemmon, E.~W., Bell, I.~H., \& McLinden, M.~O. 2022, Industrial
  \& Engineering Chemistry Research, 61, 15449,
  \dodoi{10.1021/acs.iecr.2c01427}

\bibitem[{Innes {et~al.}(2023)Innes, Tsai, \& Pierrehumbert}]{Innes_2023}
Innes, H., Tsai, S.-M., \& Pierrehumbert, R.~T. 2023, The Astrophysical
  Journal, 953, 168, \dodoi{10.3847/1538-4357/ace346}

\bibitem[{{Kite} {et~al.}(2020){Kite}, {Fegley}, {Schaefer}, \&
  {Ford}}]{Kite2020}
{Kite}, E.~S., {Fegley}, Bruce, J., {Schaefer}, L., \& {Ford}, E.~B. 2020,
  \apj, 891, 111, \dodoi{10.3847/1538-4357/ab6ffb}

\bibitem[{Koschinsky {et~al.}(2008)Koschinsky, Garbe-Sch\"onberg, Sander,
  Schmidt, Gennerich, \& Strauss}]{Koschinsky_2008}
Koschinsky, A., Garbe-Sch\"onberg, D., Sander, S., {et~al.} 2008, Geology, 36,
  615, \dodoi{10.1130/g24726a.1}

\bibitem[{{Leconte} {et~al.}(2024){Leconte}, {Spiga}, {Cl{\'e}ment}, {Guerlet},
  {Selsis}, {Milcareck}, {Cavali{\'e}}, {Moreno}, {Lellouch},
  {Carri{\'o}n-Gonz{\'a}lez}, {Charnay}, \& {Lef{\`e}vre}}]{Leconte_2024}
{Leconte}, J., {Spiga}, A., {Cl{\'e}ment}, N., {et~al.} 2024, arXiv e-prints,
  arXiv:2401.06608, \dodoi{10.48550/arXiv.2401.06608}

\bibitem[{Lemmon {et~al.}(2018)Lemmon, Bell, Huber, \& McLinden}]{LEMMON-RP10}
Lemmon, E.~W., Bell, I.~H., Huber, M.~L., \& McLinden, M.~O. 2018, {NIST
  Standard Reference Database 23: Reference Fluid Thermodynamic and Transport
  Properties-REFPROP, Version 10.0, National Institute of Standards and
  Technology}, \dodoi{https://doi.org/10.18434/T4/1502528}

\bibitem[{{Madhusudhan} {et~al.}(2023{\natexlab{a}}){Madhusudhan}, {Moses},
  {Rigby}, \& {Barrier}}]{Madhu_Faraday_2023}
{Madhusudhan}, N., {Moses}, J.~I., {Rigby}, F., \& {Barrier}, E.
  2023{\natexlab{a}}, Faraday Discussions, 245, 80, \dodoi{10.1039/D3FD00075C}

\bibitem[{{Madhusudhan} {et~al.}(2020){Madhusudhan}, {Nixon}, {Welbanks},
  {Piette}, \& {Booth}}]{Madhusudhan2020}
{Madhusudhan}, N., {Nixon}, M.~C., {Welbanks}, L., {Piette}, A. A.~A., \&
  {Booth}, R.~A. 2020, \apjl, 891, L7, \dodoi{10.3847/2041-8213/ab7229}

\bibitem[{{Madhusudhan} {et~al.}(2021){Madhusudhan}, {Piette}, \&
  {Constantinou}}]{Madhu_2021}
{Madhusudhan}, N., {Piette}, A. A.~A., \& {Constantinou}, S. 2021, \apj, 918,
  1, \dodoi{10.3847/1538-4357/abfd9c}

\bibitem[{{Madhusudhan} {et~al.}(2023{\natexlab{b}}){Madhusudhan}, Sarkar,
  Constantinou, Holmberg, Piette, \& Moses}]{Madhu_Hycean_2023}
{Madhusudhan}, N., Sarkar, S., Constantinou, S., {et~al.} 2023{\natexlab{b}},
  The Astrophysical Journal Letters, 956, \dodoi{10.3847/2041-8213/acf577}

\bibitem[{May {et~al.}(2023)May, MacDonald, Bennett, Moran, Wakeford, Peacock,
  Lustig-Yaeger, Highland, Stevenson, Sing, Mayorga, Batalha, Kirk,
  López-Morales, Valenti, Alam, Alderson, Fu, Gonzalez-Quiles, Lothringer,
  Rustamkulov, \& Sotzen}]{May_2023}
May, E.~M., MacDonald, R.~J., Bennett, K.~A., {et~al.} 2023, The Astrophysical
  Journal Letters, 959, L9, \dodoi{10.3847/2041-8213/ad054f}

\bibitem[{Melwani~Daswani {et~al.}(2021)Melwani~Daswani, Vance, Mayne, \&
  Glein}]{Daswani_2021}
Melwani~Daswani, M., Vance, S.~D., Mayne, M.~J., \& Glein, C.~R. 2021,
  Geophysical Research Letters, 48, e2021GL094143,
  \dodoi{https://doi.org/10.1029/2021GL094143}

\bibitem[{{Mol Lous} {et~al.}(2024){Mol Lous}, {Mordasini}, \&
  {Helled}}]{Mol_Lous_2024}
{Mol Lous}, M., {Mordasini}, C., \& {Helled}, R. 2024, \aap, 685, A22,
  \dodoi{10.1051/0004-6361/202349039}

\bibitem[{Moran {et~al.}(2023)Moran, Stevenson, Sing, MacDonald, Kirk,
  Lustig-Yaeger, Peacock, Mayorga, Bennett, López-Morales, May, Rustamkulov,
  Valenti, Redai, Alam, Batalha, Fu, Gonzalez-Quiles, Highland, Kruse,
  Lothringer, Ceballos, Sotzen, \& Wakeford}]{Moran_2023}
Moran, S.~E., Stevenson, K.~B., Sing, D.~K., {et~al.} 2023, The Astrophysical
  Journal Letters, 948, L11, \dodoi{10.3847/2041-8213/accb9c}

\bibitem[{Mousis {et~al.}(2020)Mousis, Deleuil, Aguichine, Marcq, Naar,
  Aguirre, Brugger, \& Gonçalves}]{Mousis_2020}
Mousis, O., Deleuil, M., Aguichine, A., {et~al.} 2020, The Astrophysical
  Journal Letters, 896, L22, \dodoi{10.3847/2041-8213/ab9530}

\bibitem[{{Nixon} \& {Madhusudhan}(2021)}]{Nixon_2021}
{Nixon}, M.~C., \& {Madhusudhan}, N. 2021, \mnras, 505, 3414,
  \dodoi{10.1093/mnras/stab1500}

\bibitem[{Pierrehumbert(2023)}]{Pierrehumbert_2023}
Pierrehumbert, R.~T. 2023, The Astrophysical Journal, 944, 20,
  \dodoi{10.3847/1538-4357/acafdf}

\bibitem[{{Rigby} {et~al.}(2024){Rigby}, {Pica-Ciamarra}, {Holmberg},
  {Madhusudhan}, {Constantinou}, {Schaefer}, {Deng}, {Lee}, \&
  {Moses}}]{Rigby_2024}
{Rigby}, F.~E., {Pica-Ciamarra}, L., {Holmberg}, M., {et~al.} 2024, arXiv
  e-prints, arXiv:2409.03683, \dodoi{10.48550/arXiv.2409.03683}

\bibitem[{{Rustamkulov} {et~al.}(2023){Rustamkulov}, {Sing}, {Mukherjee},
  {May}, {Kirk}, {Schlawin}, {Line}, {Piaulet}, {Carter}, {Batalha}, {Goyal},
  {L{\'o}pez-Morales}, {Lothringer}, {MacDonald}, {Moran}, {Stevenson},
  {Wakeford}, {Espinoza}, {Bean}, {Batalha}, {Benneke}, {Berta-Thompson},
  {Crossfield}, {Gao}, {Kreidberg}, {Powell}, {Cubillos}, {Gibson}, {Leconte},
  {Molaverdikhani}, {Nikolov}, {Parmentier}, {Roy}, {Taylor}, {Turner},
  {Wheatley}, {Aggarwal}, {Ahrer}, {Alam}, {Alderson}, {Allen}, {Banerjee},
  {Barat}, {Barrado}, {Barstow}, {Bell}, {Blecic}, {Brande}, {Casewell},
  {Changeat}, {Chubb}, {Crouzet}, {Daylan}, {Decin}, {D{\'e}sert},
  {Mikal-Evans}, {Feinstein}, {Flagg}, {Fortney}, {Harrington}, {Heng}, {Hong},
  {Hu}, {Iro}, {Kataria}, {Kempton}, {Krick}, {Lendl}, {Lillo-Box}, {Louca},
  {Lustig-Yaeger}, {Mancini}, {Mansfield}, {Mayne}, {Miguel}, {Morello},
  {Ohno}, {Palle}, {Petit dit de la Roche}, {Rackham}, {Radica},
  {Ramos-Rosado}, {Redfield}, {Rogers}, {Shkolnik}, {Southworth}, {Teske},
  {Tremblin}, {Tucker}, {Venot}, {Waalkes}, {Welbanks}, {Zhang}, \&
  {Zieba}}]{Rustamkulov_2023}
{Rustamkulov}, Z., {Sing}, D.~K., {Mukherjee}, S., {et~al.} 2023, \nat, 614,
  659, \dodoi{10.1038/s41586-022-05677-y}

\bibitem[{{Scheibe} {et~al.}(2019){Scheibe}, {Nettelmann}, \&
  {Redmer}}]{Schiebe_2019}
{Scheibe}, L., {Nettelmann}, N., \& {Redmer}, R. 2019, \aap, 632, A70,
  \dodoi{10.1051/0004-6361/201936378}

\bibitem[{{Schlichting} \& {Young}(2022)}]{Schlichting2022}
{Schlichting}, H.~E., \& {Young}, E.~D. 2022, \psj, 3, 127,
  \dodoi{10.3847/PSJ/ac68e6}

\bibitem[{{Schneider} \& {Bitsch}(2021)}]{Schneider2021a}
{Schneider}, A.~D., \& {Bitsch}, B. 2021, \aap, 654, A71,
  \dodoi{10.1051/0004-6361/202039640}

\bibitem[{Schulze-Makuch {et~al.}(2017)Schulze-Makuch, Airo, \&
  Schirmack}]{Schulze-Makuch_2017}
Schulze-Makuch, D., Airo, A., \& Schirmack, J. 2017, Frontiers in Microbiology,
  8, \dodoi{10.3389/fmicb.2017.02011}

\bibitem[{Seager {et~al.}(2021)Seager, Petkowski, G\"unther, Bains,
  Mikal-Evans, \& Deming}]{Seager_2021}
Seager, S., Petkowski, J.~J., G\"unther, M.~N., {et~al.} 2021, Universe, 7,
  \dodoi{10.3390/universe7060172}

\bibitem[{Seward \& Franck(1981)}]{Seward_1981}
Seward, T.~M., \& Franck, E.~U. 1981, Berichte der Bunsengesellschaft f\"ur
  physikalische Chemie, 85, 2, \dodoi{https://doi.org/10.1002/bbpc.19810850103}

\bibitem[{Shock \& Helgeson(1988)}]{Shock_1988}
Shock, E.~L., \& Helgeson, H.~C. 1988, Geochimica et Cosmochimica Acta, 52,
  2009.
\newblock \url{https://api.semanticscholar.org/CorpusID:97463455}

\bibitem[{Shock \& Helgeson(1990)}]{SHOCK_HELGESON_1990}
---. 1990, Geochimica et Cosmochimica Acta, 54, 915,
  \dodoi{https://doi.org/10.1016/0016-7037(90)90429-O}

\bibitem[{Shorttle {et~al.}(2024)Shorttle, Jordan, Nicholls, Lichtenberg, \&
  Bower}]{Shorttle_2024}
Shorttle, O., Jordan, S., Nicholls, H., Lichtenberg, T., \& Bower, D.~J. 2024,
  The Astrophysical Journal Letters, 962, \dodoi{10.3847/2041-8213/ad206e}

\bibitem[{{Sotin} {et~al.}(2007){Sotin}, {Grasset}, \& {Mocquet}}]{Sotin2007}
{Sotin}, C., {Grasset}, O., \& {Mocquet}, A. 2007, \icarus, 191, 337,
  \dodoi{10.1016/j.icarus.2007.04.006}

\bibitem[{{Soubiran} \& {Militzer}(2015)}]{soubiran2015}
{Soubiran}, F., \& {Militzer}, B. 2015, High Energy Density Physics, 17, 157,
  \dodoi{10.1016/j.hedp.2014.10.005}

\bibitem[{Sverjensky(2019)}]{Sverjensky_2019}
Sverjensky, D.~A. 2019, Journal of the Geological Society, 176, 348,
  \dodoi{10.1144/jgs2018-105}

\bibitem[{Sverjensky {et~al.}(2014)Sverjensky, Harrison, \&
  Azzolini}]{SVERJENSKY2014125}
Sverjensky, D.~A., Harrison, B., \& Azzolini, D. 2014, Geochimica et
  Cosmochimica Acta, 129, 125,
  \dodoi{https://doi.org/10.1016/j.gca.2013.12.019}

\bibitem[{Truong {et~al.}(2024)Truong, Glein, \& Lunine}]{Truong_2024}
Truong, N., Glein, C.~R., \& Lunine, J.~I. 2024, The Astrophysical Journal,
  976, 14, \dodoi{10.3847/1538-4357/ad7a65}

\bibitem[{Tsai {et~al.}(2021)Tsai, Innes, Lichtenberg, Taylor, Malik, Chubb, \&
  Pierrehumbert}]{Tsai_2021}
Tsai, S.-M., Innes, H., Lichtenberg, T., {et~al.} 2021, The Astrophysical
  Journal Letters, 922, \dodoi{10.3847/2041-8213/ac399a}

\bibitem[{{Tsai} {et~al.}(2024){Tsai}, {Innes}, {Wogan}, \&
  {Schwieterman}}]{Tsai_2024}
{Tsai}, S.-M., {Innes}, H., {Wogan}, N.~F., \& {Schwieterman}, E.~W. 2024,
  \apjl, 966, L24, \dodoi{10.3847/2041-8213/ad3801}

\bibitem[{{Von Damm} {et~al.}(2003){Von Damm}, Lilley, Shanks, Brockington,
  Bray, O’Grady, Olson, Graham, \& Proskurowski}]{VONDAMM_2003}
{Von Damm}, K., Lilley, M., Shanks, W., {et~al.} 2003, Earth and Planetary
  Science Letters, 206, 365,
  \dodoi{https://doi.org/10.1016/S0012-821X(02)01081-6}

\bibitem[{Wagner \& Pru\ss(2002)}]{Wagner_2002}
Wagner, W., \& Pru\ss, A. 2002, Journal of Physical and Chemical Reference
  Data, 31, 387, \dodoi{10.1063/1.1461829}

\bibitem[{Waite {et~al.}(2017)Waite, Glein, Perryman, Teolis, Magee, Miller,
  Grimes, Perry, Miller, Bouquet, Lunine, Brockwell, \& Bolton}]{Waite_2017}
Waite, J.~H., Glein, C.~R., Perryman, R.~S., {et~al.} 2017, Science, 356, 155,
  \dodoi{10.1126/science.aai8703}

\bibitem[{Wei \& Iglesia(2004)}]{WEI_2004}
Wei, J., \& Iglesia, E. 2004, Journal of Catalysis, 224, 370,
  \dodoi{https://doi.org/10.1016/j.jcat.2004.02.032}

\bibitem[{Winn \& Fabrycky(2015)}]{Winn_2015}
Winn, J.~N., \& Fabrycky, D.~C. 2015, Annual Review of Astronomy and
  Astrophysics, 53, 409, \dodoi{10.1146/annurev-astro-082214-122246}

\bibitem[{Wogan {et~al.}(2024)Wogan, Batalha, Zahnle, Krissansen-Totton, Tsai,
  \& Hu}]{wogan2024}
Wogan, N.~F., Batalha, N.~E., Zahnle, K.~J., {et~al.} 2024, The Astrophysical
  Journal Letters, 963, L7, \dodoi{10.3847/2041-8213/ad2616}

\bibitem[{{Yang} \& {Hu}(2024)}]{Yang_2024}
{Yang}, J., \& {Hu}, R. 2024, arXiv e-prints, arXiv:2406.01955,
  \dodoi{10.48550/arXiv.2406.01955}

\bibitem[{Yu {et~al.}(2021)Yu, Moses, Fortney, \& Zhang}]{Yu_2021}
Yu, X., Moses, J.~I., Fortney, J.~J., \& Zhang, X. 2021, The Astrophysical
  Journal, 914, 38, \dodoi{10.3847/1538-4357/abfdc7}

\bibitem[{Zhang \& Duan(2005)}]{ZHANG_2005}
Zhang, Z., \& Duan, Z. 2005, Physics of the Earth and Planetary Interiors, 149,
  335, \dodoi{https://doi.org/10.1016/j.pepi.2004.11.003}

\end{thebibliography}

\end{document}